\documentclass[pra,letterpaper,showpacs,showkeys]{revtex4}

\usepackage{graphicx}

\usepackage{amsthm}
\usepackage{amsmath}
\usepackage{amsfonts}
\usepackage{amscd}

\usepackage{bm}

\newtheorem{Theorem}{Theorem}[section]
\newtheorem{Def}[Theorem]{Definition}
\newtheorem{Lemma}[Theorem]{Lemma}

\begin{document}


\newcommand{\Ket}[1]{\ensuremath{\left | #1 \right \rangle}}
\newcommand{\Bra}[1]{\ensuremath{\left \langle #1 \right |}}
\newcommand{\BraKet}[2]{\ensuremath{\left \langle #1 \right |
\left. #2 \right \rangle}}
\newcommand{\Tr}[1]{\ensuremath{\mbox{Tr} \left ( #1 \right )}}
\newcommand{\PTr}[2]{\ensuremath{\mbox{Tr}_{#1} \left ( #2 \right )}}


\setlength{\unitlength}{1cm}


\title{Quantum Dynamics as an analog of Conditional Probability} 
\author{M. S. Leifer}
\affiliation{Perimeter Institute for Theoretical Physics, 31 Caroline Street North, Waterloo, Ontario, Canada, N2L 2Y5}
\email{mleifer@perimeterinstitute.ca}
\date{June 14, 2006}


\begin{abstract} 
Quantum theory can be regarded as a 
non-commutative generalization of classical probability.  From this point 
of view, one expects quantum dynamics to be analogous to 
classical conditional probabilities.  In this paper, a 
variant of the well-known isomorphism between completely positive 
maps and bipartite density operators is derived, which makes this connection 
much more explicit.  The new isomorphism is given an operational interpretation in terms of statistical correlations between ensemble preparation procedures and outcomes of measurements. Finally, the isomorphism is applied to 
elucidate the connection between no-cloning/no-broadcasting 
theorems and the monogamy of entanglement, and a simplified proof of
the no-broadcasting theorem is obtained as a byproduct.  
\end{abstract}

\maketitle


\section{Introduction}

\label{Intro}

Quantum theory can be regarded as a noncommutative generalization of classical probability theory, in which density operators play the role of probability distributions and the Cartesian product of probability spaces becomes the tensor product of Hilbert spaces (or more generally of $C^*$ algebras).  This point of view has been highly influential in the developing field of quantum information theory \cite{NielChuang}, which studies the same questions that arise in classical information theory in the noncommutative context.

However, quantum theory, as it is usually formulated, is not directly
analogous to abstract probability theory in the sense of Kolmogorv \cite{Kolmogorov29},
but is much closer to the theory of stochastic processes \cite{Doob42}.  In
nonrelativistic quantum mechanics, a quantum state is conceived as the
state of a number of subsystems at a particular time, and states at
different times are related by dynamics, generally represented as a
Completely Positive (CP) map.  In the relativistic case, there are
many such descriptions corresponding to different inertial frames,
related to each other via unitary transformations.  Nevertheless, the
states are always defined on spacelike hyperplanes, so the underlying
causal structure is still present in all of these descriptions.  This type of theory is closely analogous to a classical stochastic process, in which a state is a probability distribution over a set of random variables representing the properties of a system at a given time, and the states at different times are related by dynamics, given by a stochastic transition matrix.  

In contrast, abstract probability spaces make no assumptions about the causal structure of the events on which probabilities are defined.  Two disjoint events might refer to properties of two different subsystems at a given time, or they might refer to properties of the same subsystem at two different times.  In full generality, events need have no interpretation in terms of causal structure at all.  It is interesting ask whether quantum theory can be reformulated as an abstract noncommutative probability theory in this sense.  A first step along this road is to ask whether correlations between different subsystems and correlations between the same system before and after the application of a CP-map can be expressed using an identical formalism.  In the analogous classical case, both can be handled by conditional probabilities, so we are really asking whether a good quantum analog of conditional probability exists.  

In this paper, the question is answered in the affirmative by deriving
a variant of the isomorphism between bipartite states and CP maps
discovered by Jamio{\l}kowski \cite{Jam72} and Choi \cite{Choi75}, that makes the connection to conditional probability much more explicit.  An operational interpretation of the new isomorphism is given by showing that the same sets of correlations can be obtained in each of the two cases.

This result is interesting from the point of view of quantum
information, since many relationships have already been discovered
between the properties of bipartite quantum states and those of noisy
quantum channels \cite{VerVer03, ArrPat04}, i.e. trace preserving CP maps.  Some of these can be
extended using the new approach.  In particular, it is shown that the
various types of no-cloning/no-broadcasting theorem \cite{WZClone82,
  DieksClone82, NoBroad96} can be associated directly to statements
about the monogamy of entanglement for tripartite states
\cite{Terhal03}, i.e. the fact that if two subsystems are in a pure
entangled state, neither of them can be entangled with any other
subsystems.  As a byproduct of this, a simplified proof of the
no-broadcasting theorem is obtained. 

\subsection{Prior Work}

\label{Intro:Prior}

The central question addressed in this paper was originally raised by Ohya
\cite{Ohya83-1, Ohya83-2}.  Griffith \cite{Griffiths05}
suggested that the Jamio{\l}kowski isomorphism might be extended by
allowing a CP-map to act on a more bipartite state.  The
suggestion was not pursued in that work, but Lo Presti and d'Ariano
later developed it in the context of quantum process tomography
\cite{LoPresti03}. The specific isomorphism developed here was very much
inspired by some observations made by Fuchs \cite{Fuchs02, Fuchs03}.  During the preparation of
this manuscript, I became aware of work by Asorey
et. al. \cite{AsoreyEtAl05}, where a similar isomorphism to the one
developed here is considered.  The main novelties of the
present work are the operational interpretation of the isomorphism, and
the application to no-cloning/no-broadcasting theorems.  Also, the
case of density operators that are not of full rank is treated more
carefully here.  Finally, Cerf and Adami have developed a different
notion of quantum conditional probability \cite{CerfAdami97,
  CerfAdami98, CerfAdami99}, based on the definition of the
conditional von Neumann entropy.  

\subsection{Outline}

\label{Intro:Outline}

The remainder of this paper is structured as follows.  As preparation
for the new quantum isomorphism, the different causal structures that can give rise
to the same classical joint probabilities are reviewed in
\S\ref{Classical}.  The standard version of the Jamio{\l}kowski
isomorphism is reviewed in \S\ref{Jamiol}, and \S\ref{New} combines
the ideas of \S\ref{Classical} and \S\ref{Jamiol} to obtain the
new isomorphism for the quantum case.  \S\ref{Cloning} develops 
the application of the new isomorphism to the connection between
cloning/broadcasting and the monogamy of entanglement.
Finally, further potential applications and open questions suggested by this
work are discussed in \S\ref{Discuss}.

\section{Causal Relations and Classical Joint Probabilities}

\label{Classical}

Given two integer-valued random variables $X$ and $Y$, with joint probability
distribution $P(X = i,Y = j)$, the marginal probability distributions for $X$
and $Y$ are given by
\begin{equation}
\label{Classical:Marginal2}
\forall j \,\,\, P(X=j) = \sum_k P(X=j,Y=k) \qquad \forall k \,\,\, P(Y=k) = \sum_j P(X=j,Y=k).
\end{equation}

As is conventional in probability theory, the notation $P(X)$ is used as a stand in for $P(X=j)$, when $j$ is an arbitrary unspecified integer.  Similar definitions apply for $P(Y)$ and $P(X,Y)$.  When a random variable appears as a free index in an equation involving probabilities, then it is implicit that the equation holds for all possible values that the variable can take, and $\sum_X$ is an instruction to sum over the possible values of $X$.  With these conventions, eq. (\ref{Classical:Marginal}) may be simplified to
\begin{equation}
\label{Classical:Marginal}
P(X) = \sum_Y P(X,Y) \qquad P(Y) = \sum_X P(X,Y),
\end{equation}
and the conditional probability of $Y$, given $X$, is defined as
\begin{equation}
\label{Classical:Conditional}
P(Y|X) = \frac{P(X,Y)}{P(X)},
\end{equation}
for all values of $X$ such that $P(X) \neq 0$, and is undefined
whenever $P(X) = 0$.  Clearly,
\begin{equation}
\label{Classical:RConditional}
P(X,Y) = P(Y|X)P(X)
\end{equation}
whenever the right hand side is defined, and $P(X,Y) = 0$ otherwise.

Note that, so far no mention has been made of how the correlations
between $X$ and $Y$ arise.  $X$ and $Y$ might refer to the same physical quantity at two
different times, $Y$ differing from $X$ due to the dynamics of the
system, or they might refer to quantities associated with distinct
physical systems at the same time.  Indeed, they may have no
interpretation in terms of physical quantities at all.  In other words,
classical probability theory does not depend in any way on the causal
ordering of variables, and in particular it
does not depend on how or even whether the random variables are embedded in spacetime.
Of course, if $X$ and $Y$ are given physical meaning, then this is likely to severely
constrain the possible assignments of $P(X,Y)$ we are likely to
entertain, but this happens at the level of the \emph{application} of
probability theory to physics, and not within the abstract theory itself.

\begin{figure}
\begin{center}
\begin{picture}(15,5)
\put(0,0){\includegraphics[angle=0, width = 15cm]{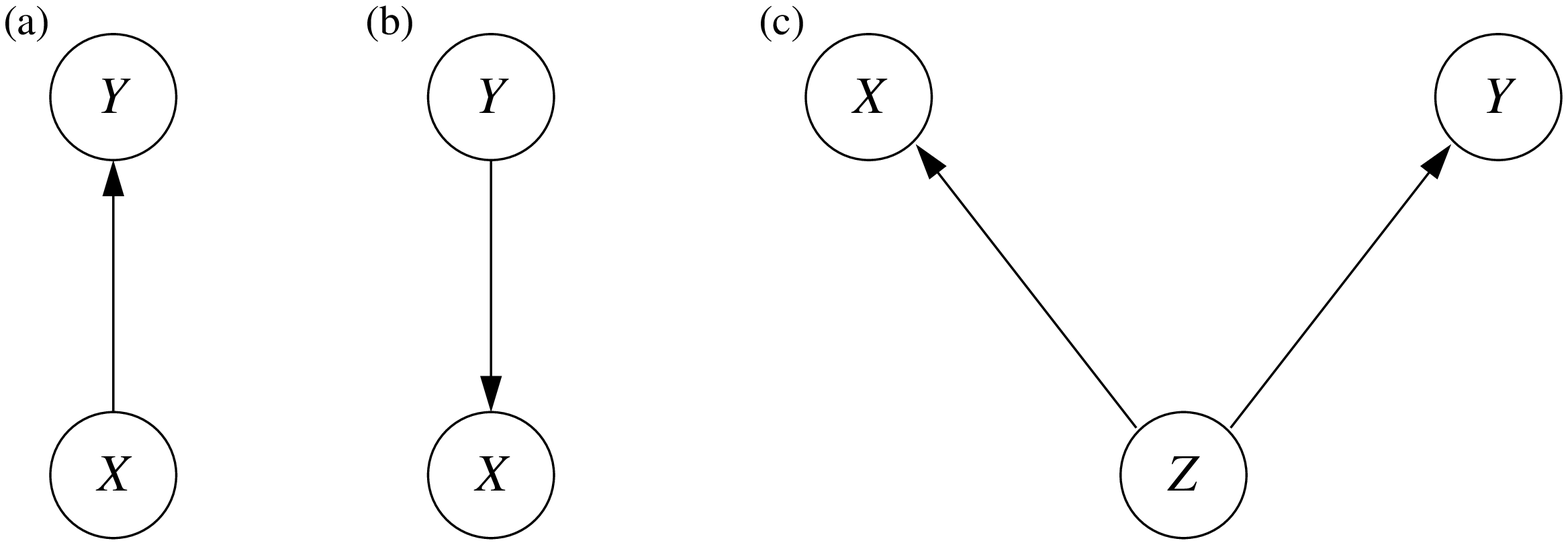}}
\put(0.2,2.5){$\Gamma_{Y|X}$}
\put(3.8,2.5){$\Gamma_{X|Y}$}
\put(8.5,2.5){$\Gamma_{X|Z}$}
\put(11.8,2.5){$\Gamma_{Y|Z}$}
\end{picture}
\end{center}
\caption{Distinct ways in which a general joint probability
  distribution $P(X,Y)$ 
  may arise. (a) $X$ is the cause of $Y$.  The generation of $Y$ must
  be in the temporal future of the generation of $X$.  For example,
  $Y$ may be the result of sending $X$ through a noisy channel
  described by a stochastic matrix $\Gamma_{Y|X}$. (b) $Y$ is
  the cause of $X$.  The generation of $X$ must be in the temporal
  future of the generation of $Y$.  For example, $X$ may be the result
  of sending $Y$ through a noisy channel described by a stochastic
  matrix $\Gamma_{X|Y}$. (c) X and Y are the result of some
  common cause, described by a random variable $Z$.  They may be
  observed at spacelike separation from one another, provided the
  points where this happens are both in the forward lightcone of the
  point where $Z$ was generated.}
\label{Classical:Distinct}
\end{figure}

In contrast, the quantum formalism handles correlations
in very different ways depending on how they arise.  Joint states of two
subsystems are handled by taking the tensor product of the underlying
Hilbert spaces, whereas correlations between the same physical
quantity at differing times are not.  This
is a weak point in the analogy between quantum
theory and classical probability, since the former cannot be viewed as
a completely abstract theory that is independent of how the
observables we are interested in are embedded in spacetime. The isomorphism of \S\ref{New} is intended to remove this deficiency,
but before moving on to the quantum case, it is helpful to understand
the different ways in which joint probability distribution $P(X,Y)$ may be described in distinct causal scenarios, the simplest of which are
depicted in fig. \ref{Classical:Distinct}.

A possible situation in which case (c) might arise is if $X$ and $Y$
represent the values of some physical quantity, associated with two distinct subsystems.
$Z$ may then represent the state of a source, which produces the two
subsystems and sends them flying out in opposite directions.  In this
situation, the values of $X$ and $Y$ could potentially be observed at
spacelike separation from one another. $P(X,Y)$ then represents the
joint state of the two subsystems, and the marginals $P(X)$ and $P(Y)$
represent their reduced states.  This is entirely
analogous to the quantum description of the joint state of two 
subsystems by a density matrix $\rho_{AB} \in \mathfrak{L}(\mathcal{H}_A
\otimes \mathcal{H}_B)$, and the descriptions of the
reduced states of the two subsystems by the reduced density matrices
$\rho_A = \PTr{B}{\rho_{AB}}$, $\rho_B = \PTr{A}{\rho_{AB}}$, where
$\mathcal{H}_A$ and $\mathcal{H}_B$ are the
Hilbert spaces associated to two subsystems $A$ and $B$, and
$\mathfrak{L}(\mathcal{H})$ denotes the space of linear operators on a
Hilbert space $\mathcal{H}$. 

A possible situation in which case (a) or (b) might arise is if $X$
and $Y$ represent the values of the same physical quantity, associated
to the same physical system at two different times $t_1 < t_2$.  In
case (a), $X$ is the value of the quantity at $t_1$ and $Y$ is its
value at $t_2$.  The transition from $X$ to $Y$ is the result of
the dynamics of the system, which may include a stochastic component
due to random external influences or a lack of knowledge about the
precise details of a deterministic dynamics.  A general dynamics is
therefore described by a stochastic matrix $\Gamma_{Y|X}$,
where $(\Gamma_{Y|X})_{ij}$ is the probability of a
transition from the state $X = j$ at $t_1$ to the state $Y = i$ at
$t_2$. The general picture we obtain from this is that the state $P(X)$ is
prepared at time $t_1$, then the dynamics $\Gamma_{Y|X}$
occurs, resulting in a final state $P(Y)$ at $t_2$. This is
summarized by the dynamical rule
\begin{equation}
\label{Classical:Evolution}
P(Y = i) = \sum_j (\Gamma_{Y|X})_{ij} P(X = j).
\end{equation}

In the quantum case, the analog of $\Gamma_{Y|X}$ is a Trace Preserving Completely Positive (TPCP) map $\mathcal{E}_{B|A} : \mathfrak{L}(\mathcal{H}_A) \rightarrow \mathfrak{L}(\mathcal{H}_B)$, which can be used to describe the
dynamics of a system that is interacting with its environment, or when
the dynamics is controlled by a random classical parameter (see
\cite{NielChuang} for further details). In this case, a density operator
$\rho_A$ is prepared at $t_1$ and then the system is subjected to a dynamical evolution according to
the TPCP map $\mathcal{E}_{B|A}$ to obtain a density
operator $\rho_B = \mathcal{E}_{B|A}(\rho_A)$ at $t_2$.  Classically, there is no reason not to consider the
two-time joint probability distribution $P(X,Y)$ that results from
combining the preparation $P(X)$ with the dynamics $\Gamma_{Y|X}$.  To do this, we need only
define the conditional distributions $P(Y|X)$, since the joint is then
given by eq. (\ref{Classical:RConditional}).  Comparing
eq. (\ref{Classical:Evolution}) with eqs. (\ref{Classical:Marginal})
and (\ref{Classical:RConditional}), we see that setting
\begin{equation}
 P(Y = i|X = j) = (\Gamma_{Y|X})_{ij},
\end{equation} 
for all $i,j$ such that $P(X = j) \neq 0$ gives the desired result. Note that, for a fixed preparation $P(X)$, we may vary the dynamics
arbitrarily for all values of $X$ that have no support in $P(X)$, without
affecting the conditional distribution $P(Y|X)$, or the joint $P(X,Y)$.  Conversely, knowing
$P(Y|X)$ or $P(X,Y)$ only specifies the dynamics on the support of $P(X)$.

Now, the set of joint probability distributions obtainable in cases
(a) and (c) are precisely the same, so we can define an isomorphism
between the pair of objects consisting of a preparation and a dynamics
and the joint state of two subsystems
\begin{equation}
\label{Classical:NewIso}
(P(X), \Gamma_{Y|X}^r) \leftrightarrow P(X,Y).
\end{equation} 
Here, $\Gamma_{Y|X}^r$ refers to the restriction of the
dynamics $\Gamma_{Y|X}$ to the support of $P(X)$, and is in one-to-one
correspondence with the conditional probability $P(Y|X)$. The left
hand side of eq. (\ref{Classical:NewIso}) can be thought of as a
description of a case (a) scenario and the right hand side as a
description of a case (c) scenario.
This may seem like an unnecessarily complicated restatement of what is
essentially the definition of conditional probability, but it is worth
remarking upon because the new isomorphism of \S\ref{New} is 
the quantum analog of this.  That is, we construct an isomorphism
between the pair of objects consisting of a preparation and a dynamics,
and the joint state of two subsystems:
\begin{equation}
\label{Classical:NewPreview}
(\rho_A, \mathcal{E}_{B|A}^r) \leftrightarrow \tau_{AB},
\end{equation}
where $\mathcal{E}_{B|A}^r$ denotes the restriction of a
TPCP map $\mathcal{E}_{B|A}$ to the support of $\rho_A$.  The object
$\mathcal{E}^r_{B|A}$ is to be thought of as a quantum analog of
conditional probability, playing the same role as $\Gamma_{Y|X}^r$
does in classical probability theory.

\section{The Jamio{\l}kowski Isomorphism}

\label{Jamiol}

In this section, the standard Jamio{\l}kowski isomorphism
is reviewed.  This relates CP-maps $\mathcal{E}_{B|A}$ to bipartite
states $\tau_{AB}$, without introducing the state $\rho_A$ that
appears in eq. (\ref{Classical:NewPreview}), and is later shown
to be a special case of the more general isomorphism described in
\S\ref{New}.  Sections \S\ref{Jamiol:Pure} and \S\ref{Jamiol:Mixed}
give the mathematical statement of the isomorphism and
\S\ref{Jamiol:Physics} gives its operational interpretation.  Comments
about the isomorphism that will be important in what 
follows are made in \S\ref{Jamiol:Facts}.  The discussion is intended
to be self contained, but the interested reader can find detailed
overviews different aspects of the isomorphism in \cite{ArrPat04, ZyczBeng04}. 

\subsection{Operators and Pure States}

\label{Jamiol:Pure}

Let $\mathcal{H}_A$, $\mathcal{H}_B$ be Hilbert spaces of 
dimension $d_A$ and $d_B$ respectively and let 
$\{\Ket{j}_A\}$ be an orthonormal basis for $\mathcal{H}_A$. An 
operator $R_{B|A}: \mathcal{H}_A \rightarrow
\mathcal{H}_B$ is isomorphic to a (generally unnormalized) pure state
$\Ket{\Psi_R}_{AB} \in \mathcal{H}_A \otimes \mathcal{H}_B$ given by  
\begin{equation} \Ket{\Psi_R}_{AB} = 
\frac{1}{\sqrt{d_A}}\sum_{j=1}^{d_A} \Ket{j}_A \otimes R_{B|A'} \Ket{j}_{A'} = 
I_A \otimes R_{B|A'} \Ket{\Phi^+}_{AA'}, 
\end{equation} 
where $A'$ denotes an additional system with the same Hilbert space as
$A$ \footnote{Generally, $X', X'', \ldots$ are
used as labels for ancillary systems with the same Hilbert space as
the system labeled by $X$, and $\mathcal{H}_{X'}, \mathcal{H}_{X''},
\ldots$ are synonyms for $\mathcal{H}_X$.}, $\Ket{\Phi^+}_{AA'} =
\frac{1}{\sqrt{d_A}} \sum_{j=1}^{d_A} \Ket{j}_A\otimes\Ket{j}_{A'}$ is
a maximally entangled state on $\mathcal{H}_A \otimes
\mathcal{H}_{A'}$ and $I_A$ is the identity operator on
$\mathcal{H}_A$ \footnote{Generally, $I_X$ denotes the identity
  operator on $\mathcal{H}_X$ for an arbitrary system label $X$.}. 
 
To see that this is an isomorphism, note that the action of $R_{B|A}$ on a pure state $\Ket{\psi}_A$ can be recovered from
  $\Ket{\Psi_R}_{AB}$ via  
\begin{equation} 
\label{Jamiol:Pure:Reverse}
R_{B|A} \Ket{\psi}_A = d_A \Bra{\Phi^+}_{AA'}
\Ket{\psi}_A \otimes \Ket{\Psi_R}_{A'B}. 
\end{equation}

\subsection{Completely Positive Maps and Mixed States}
\label{Jamiol:Mixed}

The isomorphism can be extended from operators to CP maps, $\mathcal{E}_{B|A}: \mathfrak{L}(\mathcal{H}_A)\rightarrow \mathfrak{L}(\mathcal{H}_B)$, where $\mathfrak{L}(\mathcal{H})$ 
is the space of linear operators on a Hilbert space 
$\mathcal{H}$.   An arbitrary CP map can be
characterized by a set of linear operators $R^{(\mu)}_{B|A} : \mathcal{H}_A \rightarrow \mathcal{H}_B$, known as Kraus
operators. The action of $\mathcal{E}_{B|A}$ on density 
operators $\rho_A \in \mathfrak{L}(\mathcal{H}_A)$ is given by 
\begin{equation} 
\mathcal{E}_{B|A} ( \rho_A ) = \sum_\mu R^{(\mu)}_{B|A} \rho_A R^{(\mu)\dagger}_{B|A},
\end{equation} 
where $^\dagger$ denotes the conjugate transpose.
Note that a CP map typically has more than one decomposition
into Kraus operators.
 
If $\sum_\mu R^{(\mu)\dagger}_{B|A} R^{(\mu)}_{B|A} = I_A$ then the map is called a Trace Preserving
Completely Positive (TPCP) map and it can be implemented with
certainty by introducing an ancilla, performing a unitary
transformation, and then taking the partial trace over a subsystem
(see \cite{NielChuang} for details).  
On the other hand, if  $\sum_\mu R^{(\mu)\dagger}_{B|A}
  R^{(\mu)}_{B|A} < I_A$, then the action
of the CP map gives the (unnormalized) updated state after obtaining a
particular outcome in a generalized measurement, and it cannot be implemented with
certainty.  In what follows, the main focus is on TPCP maps, but
comments on the general case are made in \S\ref{Jamiol:Facts}.  

The state isomorphic to $\mathcal{E}_{B|A}$ is generally
mixed and is given by 
\begin{align} 
\tau_{AB} & = \sum_\mu \Ket{\Psi_{R^{(\mu)}}}_{AB}
\Bra{\Psi_{R^{(\mu)}}}_{AB} \label{Jamiol:Mixed:KF} \\
& = \mathcal{I}_{A} \otimes \mathcal{E}_{B|A'} \left (
  \Ket{\Phi^+}_{AA'}\Bra{\Phi^+}_{AA'} \right ),  \label{Jamiol:Mixed:GF}
\end{align} 
where $\mathcal{I}_A$ is the identity CP map on
$\mathfrak{L}(\mathcal{H}_A)$.  Note that the state
$\tau_{AB}$ depends only on $\mathcal{E}_{B|A}$ and not on a particular decomposition into Kraus
operators.  The form in eq. (\ref{Jamiol:Mixed:KF}) gives different
pure-state decompositions of the same density operator as the Kraus
decomposition is varied and all pure state decompositions of
$\tau_{AB}$ can be obtained in this way.

The reverse direction of the isomorphism is similar to
eq. (\ref{Jamiol:Pure:Reverse}).  The action of $\mathcal{E}_{B|A}$ on an arbitrary $\sigma_A \in
\mathfrak{L}(\mathcal{H}_A)$ is given by
\begin{equation}
\label{Jamiol:Mixed:Reverse}
\mathcal{E}_{B|A} \left ( \sigma_A \right ) =
d_A^2 \Bra{\Phi^+}_{AA'} \sigma_A \otimes \tau_{A'B}
\Ket{\Phi^+}_{AA'}.
\end{equation}

\subsection{Operational Interpretation}
\label{Jamiol:Physics}

So far, the isomorphism has been stated as a mathematical fact.  For TPCP maps, it
obtains operational meaning via \emph{noisy gate teleportation}, the
obvious extension of a protocol described in \cite{NielChuang97} for
unitary gates.

Suppose Alice holds an unknown state $\sigma_A$ and that Bob wishes to
end up with the transformed state $\mathcal{E}_{B|A}(\sigma_A)$, where $\mathcal{E}_{B|A}$ is a TPCP map.
They also share a copy of the isomorphic state
$\tau_{A'B}$ and they wish to achieve the task via Local Operations and
Classical Communication (LOCC).  If the map $\mathcal{E}_{B|A}$ is just the identity $\mathcal{I}_A$, then the task can be
achieved via the usual teleportation protocol, since
$\tau_{A'B}$ is maximally entangled in this case. In general, the task
can be achieved with probability of success at least
$\frac{1}{d_A^2}$, since if Alice makes a measurement in a basis that
includes the state $\Ket{\Phi^+}_{AA'}$ then Bob will receive the
correct transformed state whenever Alice gets this outcome, as can be
seen from eq. (\ref{Jamiol:Mixed:Reverse}).  For certain special
maps, Bob can correct his
state when Alice does not get the right outcome, as in the
teleportation protocol, but this is not possible in general.

\subsection{Remarks}
\label{Jamiol:Facts}

The following facts about the Jamio{\l}kowski isomorphism are
important in what follows.  Firstly, note that the isomorphism is basis dependent, since the definition of the state $\Ket{\Phi^+}$ makes use of a particular basis.  The association between a bipartite state and a CP-map is unique, up to this choice of basis.  

Secondly, if $\mathcal{E}_{B|A}$ is a
TPCP map, then the state $\tau_{AB}$ always has the
maximally mixed state as the reduced density operator for system $A$,
i.e. $\PTr{B}{\tau_{AB}} = \frac{I_A}{d_A}$.  This can
be deduced from eq. (\ref{Jamiol:Mixed:GF}) and the fact that the state
$\Ket{\Phi^+}_{AA'}$ is maximally mixed on $A$.  To obtain an
arbitrary state via the isomorphism, one has to use the more general CP
maps that cannot be implemented deterministically.  For example, the pure
product state $\Ket{00}_{AB}$ corresponds to the projection operator
$\Ket{0}_B\Bra{0}_A$ that results from obtaining the $\Ket{0}$ outcome
of a measurement in the computational basis, and then relabeling
system $A$ to $B$.  A major difference between the standard isomorphism and the new variant
described in \S\ref{New}, is that in the new version, all bipartite states
are obtained with just TPCP maps.  

\section{A new variant of the Jamio{\l}kowski isomorphism}
\label{New}

In this section, the new isomorphism is described.  It is constructed and shown to be an an isomorphism in \S\ref{New:Construction}.  \S\ref{New:Operational} gives the operational interpretation of the isomorphism.  Finally, \S\ref{New:Commute} and \S\ref{New:Remarks} describe some properties of the isomorphism that are exploited in the applications that follow.

\subsection{Construction of the isomorphism}

\label{New:Construction}

Recall from \S\ref{Classical} that the aim is to construct an
isomorphism
\begin{equation}
(\rho_A, \mathcal{E}_{B|A}^r) \leftrightarrow \tau_{AB},
\end{equation}
where $\mathcal{E}_{B|A}^r$ denotes the restriction of a
TPCP map $\mathcal{E}_{B|A}$ to the support of $\rho_A$. 

We begin by describing the forward direction of the isomorphism, $(\rho_A, \mathcal{E}_{B|A}^r) \rightarrow \tau_{AB}$.  First, construct the state 
\begin{equation}
\label{New:Phi}
\Ket{\Phi}_{AA'} = \sqrt{d_A} \left ( \rho_A^T \right )^{\frac{1}{2}} \otimes I_{A'} \Ket{\Phi^+}_{AA'},
\end{equation}
where $^T$ denotes transpose in the basis used to define
$\Ket{\Phi^+}_{AA'}$.  Note that this is a normalized state since
$\BraKet{\Phi}{\Phi}_{AA'} = \Tr{\rho_A^T} = 1$.  Also,
$\PTr{A'}{\Ket{\Phi}\Bra{\Phi}_{AA'}} = \rho_A^T$  and $\PTr{A}{\Ket{\Phi}\Bra{\Phi}_{AA'}}$ is identical to $\rho_A$, so that the action of $\mathcal{E}_{B|A'}^r$ on system $A'$ is well defined for this state. Finally, define
\begin{equation}
\label{New:RhoAB}
\tau_{AB} = \mathcal{I}_A \otimes \mathcal{E}_{B|A'}^r \left ( \Ket{\Phi}\Bra{\Phi}_{AA'} \right ),
\end{equation}
which is a normalized state because $\mathcal{E}_{B|A'}^r$ is the restriction of a TPCP map.

For the reverse direction  $\tau_{AB} \rightarrow (\rho_A, \mathcal{E}_{A \rightarrow B}^r)$, begin by defining the state
\begin{equation}
\label{New:RhoA}
\rho_A = \tau_A^T = \PTr{B}{\tau_{AB}}^T.
\end{equation}
To define the map $\mathcal{E}_{B|A}^r$, care must be taken if
$\tau_A$ is not invertible.  If this is the case, define
$\tau_A^{-1}$ to be the inverse restricted to the support of
$\tau_A$.  This means that its nonzero eigenvalues are the
reciprocals of the nonzero eigenvalues of $\tau_A$ and they are
associated with the same eigenvectors, and that the zero eigenspaces
of $\tau_A$ and $\tau_A^{-1}$ are the same.  Now, define the state
\begin{equation}
\label{New:Sigma}
\sigma_{AB} = \tau_A^{-\frac{1}{2}} \otimes I_B \tau_{AB} \tau_A^{-\frac{1}{2}} \otimes I_B.
\end{equation}
This is a density operator with $\sigma_A = \PTr{B}{\sigma_{AB}} =
\frac{1}{d_A^r} P_A$, where $P_A$ is the projector onto the support of
$\tau_A$ and $d_A^r$ is the rank of $\tau_A$.  The associated subspace, $P_A \mathcal{H}_A$, is also a
Hilbert space, for which $P_A$ is the identity operator, so $\sigma_A$
is maximally mixed on this subspace.  Thus, $\sigma_{AB}$ is uniquely associated with a TPCP map $\mathcal{E}^r_{B|A}: \mathfrak{L}(P_A \mathcal{H}_A) \rightarrow \mathfrak{L}(H_B)$ via the standard Jamio{\l}kowski isomorphism.

\begin{figure}
\begin{center}
\begin{picture}(15,9)
\put(0,0){\includegraphics[angle=0, width = 15cm]{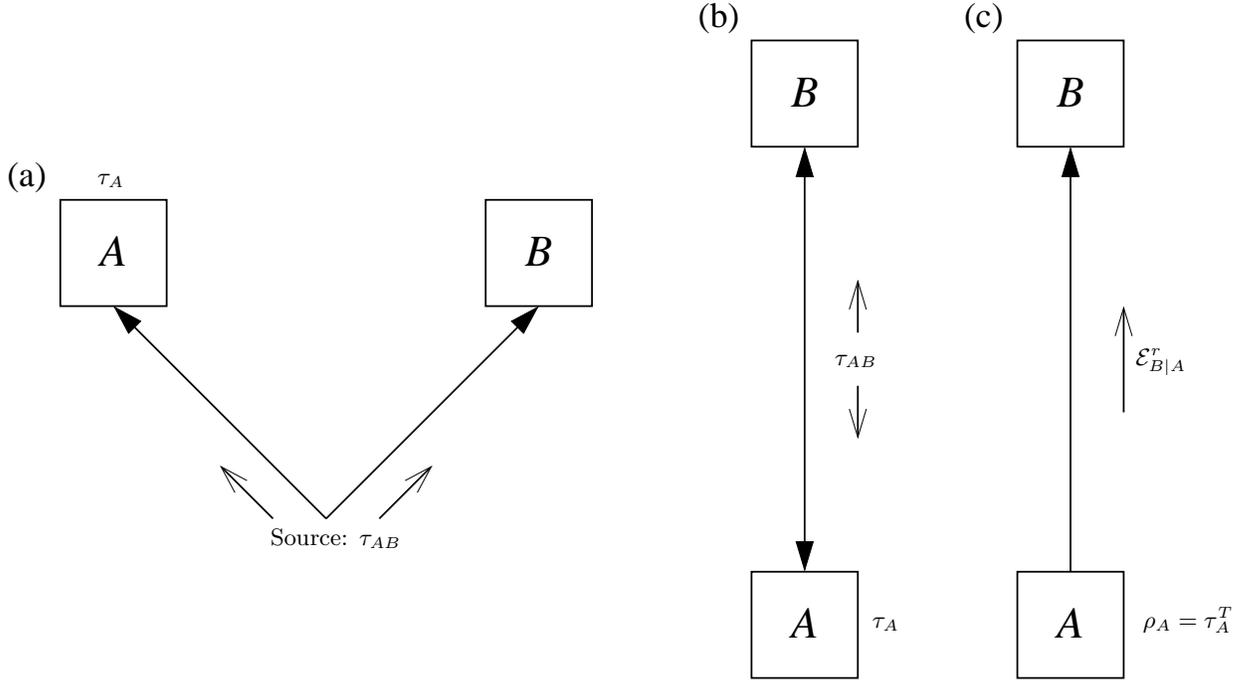}}
\put(3.5,1.8){Source: $\tau_{AB}$}
\put(1.2,6.6){$\tau_A$}
\put(11,4.2){$\tau_{AB}$}
\put(11.5,0.7){$\tau_A$}
\put(15,4.2){$\mathcal{E}_{B|A}^r$}
\put(15.1,0.7){$\rho_A = \tau_A^T$}
\end{picture}
\end{center}
\caption{\label{New:Explain}In these diagrams, time flows up the page.  Starting from (a), the space and time axes are interchanged and the diagram is ``stretched out'' to arrive at (b).  This does not describe a possible experiment, since we cannot send system $A$ backwards in the time direction.  In order to arrive at a feasible experiment, some arrows must be reversed, giving rise to (c).  The transpose on $\tau_A$ is an artifact of this time reversal.}
\end{figure}

In the above construction, eq. (\ref{New:Sigma}) can be viewed as a
direct analog of the definition of conditional probability $P(Y|X) =
P(X,Y)/P(X)$, since conjugation by $\tau_A^{-\frac{1}{2}}$ reduces to
elementwise division in the case where $\tau_{AB}$ is diagonal in an
basis $\{ \Ket{\phi_j}_A \otimes \Ket{\psi_k}_B \}$, where the
$\Ket{\phi_j}_A$ form an orthonormal basis for $\mathcal{H}_A$ and the
$\Ket{\psi_k}$ form an orthonormal basis for $\mathcal{H}_B$.  The
introduction of the transpose in eq. (\ref{New:RhoA}) is due to a
time-reversal implicit in the construction, which is illustrated by
fig. \ref{New:Explain}. Note that in the case where $\rho_A = \frac{I_A}{d_A}$, this construction reduces to the standard Jamio{\l}kowski isomorphism.

\begin{Theorem}
The construction described above is an isomorphism
\end{Theorem}
\begin{proof}
The above relations define an isomorphism if it can be shown that one
obtains the same pair $(\rho_A, \mathcal{E}_{B|A}^r)$ on applying the
forward and reverse directions in sequence.  To check this for the state $\rho_A$, let $R^{(\mu)}_{B|A'}$ be a set of Kraus operators for $\mathcal{E}^r_{B|A'}$.  Then,
\begin{equation}
\tau_{AB} = \sum_{jk\mu}   \left ( \rho_A^T \right )^{\frac{1}{2}} \Ket{j} \Bra{k}  \left ( \rho_A^T \right )^{\frac{1}{2}} \otimes R^{(\mu)}_{B|A'} \Ket{j}\Bra{k} R^{(\mu)\dagger}_{B|A'}.
\end{equation}
Taking the partial trace gives
\begin{equation}
\tau_A = \sum_{jk\mu} \left ( \rho_A^T \right )^{\frac{1}{2}} \Ket{j} \Bra{k}  \left ( \rho_A^T \right )^{\frac{1}{2}} \Bra{k} \sum_{\mu} R^{(\mu)\dagger}_{B|A'} R^{(\mu)}_{B|A'} \Ket{j}.
\end{equation}
Since $\mathcal{E}_{B|A}^r$ is trace preserving $\sum_{\mu} R^{(\mu)\dagger}_{B|A'} R^{(\mu)}_{B|A'} = I_{A'}$, so
\begin{equation}
\tau_A = \left ( \rho_A^T \right )^{\frac{1}{2}} \sum_j \Ket{j}\Bra{j} \left ( \rho_A^T \right )^{\frac{1}{2}} = \left ( \rho_A^T \right )^{\frac{1}{2}} \left ( \rho_A^T \right )^{\frac{1}{2}} = \rho_A^T,
\end{equation}
which gives the correct state for the reverse direction.

  To check the map
$\mathcal{E}_{B|A}^r$, note that the state $\Ket{\Phi^+}_{AA'}$ that
appears in eq. (\ref{New:Phi}), can be replaced
by the state $\Ket{\Phi^+}_{AA'}^r = \sqrt{\frac{d_A}{d_A^r}} P_A \otimes
I_{A'} \Ket{\Phi^+}_{AA'}$, provided $\sqrt{d_A}$ is also replaced by $\sqrt{d_A^r}$
in eq. (\ref{New:Phi}).  This is because $\left ( \rho_A^T \right )^{\frac{1}{2}}$ only has support on the subspace that $P_A$ projects onto, so the state $\Ket{\Phi}_{AA'}$ obtained will be the same. The action of $\mathcal{E}_{B|A'}^r$ is well
defined on $\Ket{\Phi^+}^r_{AA'}$ and the two steps of the construction commute,
so that the same state $\tau_{AB}$ is obtained by applying the CP-map to $\Ket{\Phi^+}_{AA'}^r$, followed
by conjugation with $\left ( \rho_A^T \right )^{\frac{1}{2}}$. The
state $\Ket{\Phi^+}_{AA'}^r$ is maximally entangled on the Hilbert
space $P_A\mathcal{H}_A \otimes P_A \mathcal{H}_A$, and so the
state $\sigma_{AB} = \mathcal{I}_A \otimes\mathcal{E}^r_{B|A'}
(\Ket{\Phi^+}^r\Bra{\Phi^+}^r_{AA'})$ is the obtained from
applying the standard Jamio{\l}kowski isomorphism to
$\mathcal{E}^r_{B|A}$.  On applying the reverse construction, the same
state $\sigma_{AB}$ is obtained in eq. (\ref{New:Sigma}), and because
states and maps are uniquely related by the standard isomorphism, the map
$\mathcal{E}_{B|A}^r$ that we started with is recovered from this procedure.
\end{proof}

\subsection{Operational Interpretation}

\label{New:Operational}

Unlike the standard Jamio{\l}kowski isomorphism, the new isomorphism does not have an
immediate operational interpretation in terms of noisy gate
teleportation.  However, there is a sense in which $\tau_{AB}$ and the
pair $( \rho_A, \mathcal{E}_{B|A}^r)$ are operationally
indistinguishable.  To understand this, we need to recall the role of Positive Operator Valued Measures (POVMs) in describing generalized quantum measurements \cite{NielChuang}, and explain their correspondence to ensemble
preparations of density operators.

A POVM is a set of positive operators that
sum to the identity.  Here, POVMs are denoted by upper-case letters
$M,N,\ldots$.  The operators within a POVM are denoted by the corresponding
boldface letter, e.g. $M = \{\bm{M}^{(m)}\}$, where the
superscript $m$ is a positive integer used to distinguish the
operators within POVM. 

POVMs are normally used to compute the probabilities for the possible outcomes of generalized measurements.  Let the possible outcomes be labeled by the same integers as the POVM elements, so that the generalized Born rule is
\begin{equation}
\label{New:Born}
P(M = m) = \Tr{\bm{M}^{(m)} \rho}.
\end{equation}
Note that the symbol $M$, which stands for a collection of operators,
is also being used to denote the random variable generated by the
measurement.  It should be clear from the context which of the two
meanings is intended.

It is convenient to extend the random variable notation used in \S\ref{Classical} to POVMs, by leaving the index $m$ implicit.  With this, the POVM is written as $M = \{\bm{M}\}$, and eq. (\ref{New:Born}) reduces to
\begin{equation}
\label{New:SimpleBorn}
P(M) = \Tr{\bm{M} \rho}.
\end{equation}

Generally, a POVM only describes the outcome statistics of a measurement and does
not specify how the state is to be updated on obtaining a particular
outcome. The update rule that should be used depends on the details of
the interaction between the system being measured and the measuring
device. Amongst the possible update rules, a particularly natural choice is
\begin{equation}
\label{New:Luders}
\rho(M) = \frac{\sqrt{\bm{M}} \rho \sqrt{\bm{M}}}{P(M)},
\end{equation}
since this reduces to the L{\"u}ders-von Neumann projection postulate
\cite{Luders} in the case where each $\bm{M}$ is a projection
operator.   If a POVM $M$ is measured on a state $\rho$, generating the
probability distribution of eq. (\ref{New:SimpleBorn}) and the state is updated according to eq. (\ref{New:Luders})
then we refer to this an \emph{$M$-measurement} of $\rho$.  The update rule for an $M$-measurement is not important for the operational interpretation developed in this section, but it is used in \S\ref{New:Commute} and in the applications of \S\ref{Cloning}. 

Although POVMs are normally used to describe measurements, they can also be used to describe the different methods of preparing a density operator $\rho$.  This is demonstrated by the following lemma.

\begin{Lemma} Let $\rho$ be a density operator and $M = \{\bm{M}\}$ a POVM.  Define $P(M) = \Tr{\bm{M} \rho}$ and let $\rho(M) = \frac{\sqrt{\rho}\bm{M}\sqrt{\rho}}{p(M)}$ whenever $P(M) > 0$.  Then, $\rho = \sum_M P(M) \rho(M)$ is an ensemble decomposition of $\rho$ into a convex combination of density operators.  Conversely, any ensemble decomposition of $\rho$ is related to a POVM in this way.
\end{Lemma}

\begin{proof}
It is clear from the definition of a POVM that $0 \leq p(M) \leq 1$ and $\sum_M p(M) = 1$.  The operators $\rho(M)$ are positive, since they are of the form $A^\dagger A$ for $A = \sqrt{\bm{M}}\sqrt{\rho}$.  They also have unit trace, so they are density operators.  Furthermore, $\sum_M P(M) \rho(M) = \sum_M \sqrt{\rho}\bm{M}\sqrt{\rho} = \rho$, by virtue of the fact that POVM operators sum to the identity.

To prove the converse, let $\rho = \sum_M P(M) \rho(M)$ be an ensemble
decomposition of $\rho$.  Then, define positive operators $\bm{M}^r =
P(M) \rho^{-\frac{1}{2}} \rho(M) \rho^{-\frac{1}{2}}$, where
$\rho^{-\frac{1}{2}}$ is the restricted inverse of
$\rho^{\frac{1}{2}}$ as defined in \S\ref{New:Construction}.  These satisfy $\Tr{\bm{M}^r\rho} = P(M)$ and
$\frac{\sqrt{\rho} \bm{M}^r \sqrt{\rho}}{p(M)} = \rho(M)$.  Clearly,
$\sum_M \bm{M}^r = P_\rho$, where $P_\rho$ is the projector onto the
support of $\rho$.  Let $P_\rho^\perp$ be the projector onto the
orthogonal complement of this subspace and choose a set positive
operators $\bm{M}^s$ of the same cardinality as $M$, supported only on
this orthogonal complement, that sum to $P_\rho^\perp$.  Since $\bm{M}^s
\rho^{1/2} = 0$, the operators $\bm{M} = \bm{M}^r + \bm{M}^s$
satisfy $\frac{\sqrt{\rho} \bm{M} \sqrt{\rho}}{p(M)} = \rho(M)$ and
sum to the identity, as required.
\end{proof}

The above lemma shows that POVMs may be used to describe ensemble
preparations of density operators as well as measurements.  For a POVM $M$, and a density operator $\rho$,
 an \emph{$M$-preparation} of $\rho$ is defined to be the procedure of generating a classical random variable with distribution $P(M) = \Tr{\bm{M}\rho}$, and then preparing the corresponding density matrix $\rho(M) =  \frac{\sqrt{\rho}\bm{M}\sqrt{\rho}}{p(M)}$.  

\begin{figure}
\begin{center}
\begin{picture}(15,12)
\put(0,0){\includegraphics[angle=0, width = 15cm]{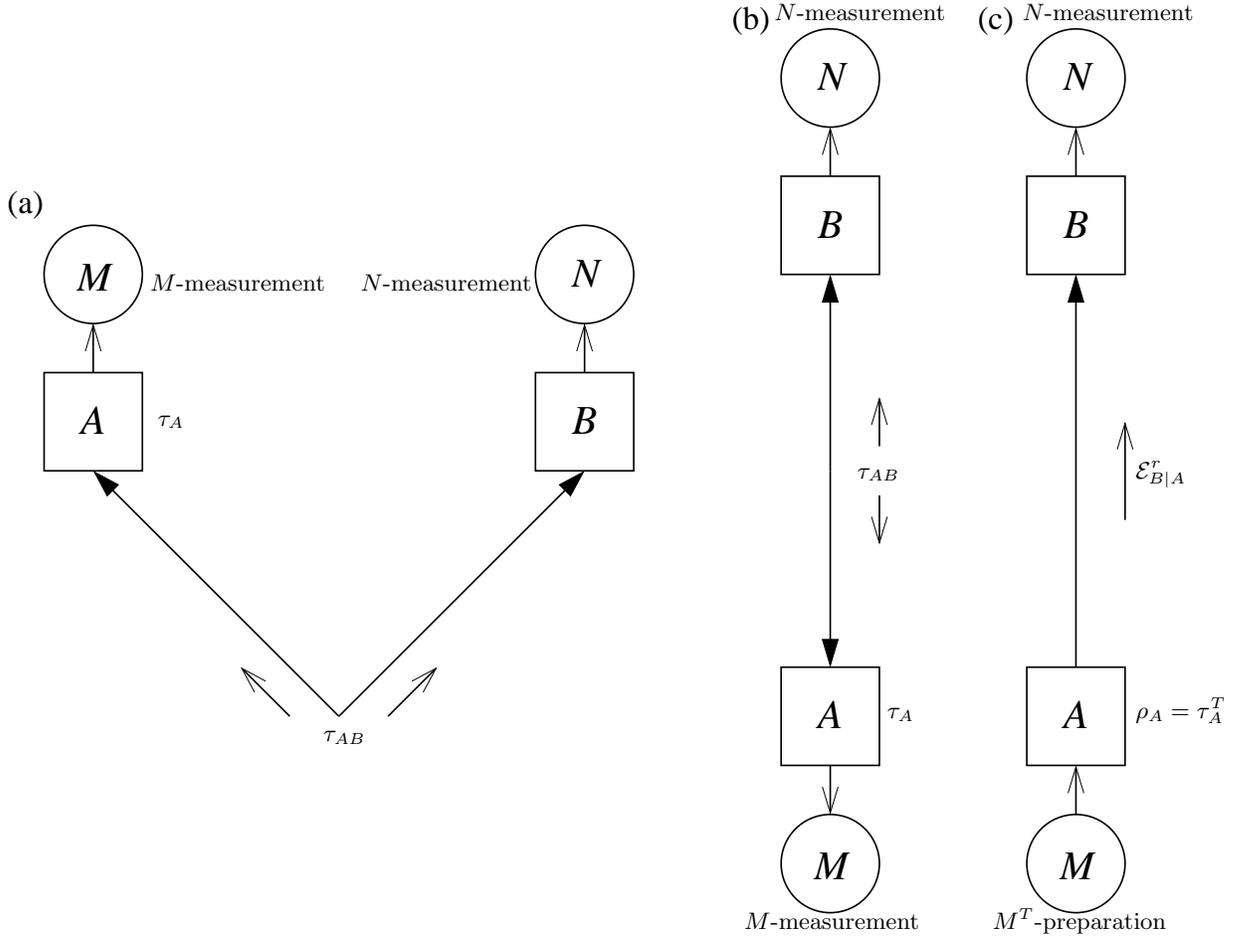}}
\put(1.9,8.3){$M$-measurement}
\put(4.7,8.3){$N$-measurement}
\put(4.2,2.3){$\tau_{AB}$}
\put(2,6.5){$\tau_A$}
\put(9.8,-0.2){$M$-measurement}
\put(11.7,2.6){$\tau_A$}
\put(11.3,5.8){$\tau_{AB}$}
\put(10.2,11.9){$N$-measurement}
\put(13.1,-0.2){$M^T$-preparation}
\put(15,2.6){$\rho_A = \tau_A^T$}
\put(15,5.8){$\mathcal{E}_{B|A}^r$}
\put(13.5,11.9){$N$-measurement}
\end{picture}
\end{center}
\caption{\label{New:Explain2}This figure represents the same experiments as fig. \ref{New:Explain}, with the addition of measurements and preparations.  (a) is obtained by simply adding $M$ and $N$-measurements to fig. \ref{New:Explain}(a).  In (b), the space and time axes have been swapped and the diagram has been ``stretched out''.  As with fig. \ref{New:Explain}(b) this does not represent a possible experiment.  To obtain a feasible experiment, in addition to the transformations of fig. \ref{New:Explain}, the $M$-measurement must be transformed into a preparation, leading to (c).  The transpose is an artifact of this time-reversal.}
\end{figure}

At this stage, the relation between the statistics obtainable from
a bipartite state $\tau_{AB}$ and those obtainable from the isomorphic pair
$(\rho_A,\mathcal{E}^r_{B|A})$ can be stated and proved.  It can be understood schematically from the idea that the isomorphism represents an interchange of space and time axes, as shown in fig. \ref{New:Explain2}.

\begin{Theorem}
Let $\tau_{AB}$ be a bipartite state that is related to a pair
$(\rho_A,\mathcal{E}_{B|A})$ by the isomorphism of
\S\ref{New:Construction}.   Let $M$ and $N$ be arbitrary POVMs, defined on $\mathcal{H}_A$ and $\mathcal{H}_B$ respectively, and let $M^T$ be the POVM obtained by taking the transpose of all the operators in $M$ with respect to the basis used to construct the isomorphism.  Then, the joint probability distribution of $M$ and $N$ measurements on $\rho_{AB}$, performed in parallel, is the same as the joint probability distribution of the sequence of operations consisting of an $M^T$-preparation of $\rho_A$, followed by evolution according to $\mathcal{E}_{B|A}$, followed by an $N$ measurement.
\end{Theorem}

\begin{proof}
The proof is by direct computation.  Let $P(M, N)  =  \Tr{\bm{M}_A
  \otimes \bm{N}_B \tau_{AB}}$ be the probability distribution of the
two measurements performed in parallel on the state $\tau_{AB}$ and
let $Q(M,N) = \PTr{B}{\bm{N}_B \mathcal{E}_{B|A}^r \left (
  \rho_A^{\frac{1}{2}} \bm{M}^T_A \rho_A^{\frac{1}{2}} \right )}$ be
the joint probability distribution of the $M^T$-preparation, followed
by evolution according to $\mathcal{E}_{B|A}^r$, followed by the
$N$-measurement.  Let $\{\Ket{j}\}$ be the basis of $\mathcal{H}_A$ in
which the isomorphism is defined. Then,
\begin{eqnarray}
P(M,N) & = & \Tr{\bm{M}_A \otimes \bm{N}_B \rho_{AB}} \\
& = & \Tr{\bm{M}_A \otimes \bm{N}_B \tau_A^{\frac{1}{2}} \otimes I_B \sigma_{AB} \tau_A^{\frac{1}{2}} \otimes I_B},
\end{eqnarray}
where $\sigma_{AB}$ is the state defined in eq. (\ref{New:Sigma}).  Since $\sigma_{AB} = \mathcal{I}_A \otimes \mathcal{E}^r_{B|A'} \left ( \Ket{\Phi^+}^r\Bra{\Phi^+}^r_{AA'} \right )$, this gives
\begin{equation}
P(M,N)  =  \Tr{\tau_A^{\frac{1}{2}} \bm{M}_A \tau_A^{\frac{1}{2}} \otimes \bm{N}_B \frac{1}{d_A^r} \sum_{j,k = 1}^{d_A^r} \Ket{j}\Bra{k}_A \otimes \mathcal{E}_{B|A}^r \left ( \Ket{j}\Bra{k}_{A} \right )}.
\end{equation}
Let $\{R_{B|A}^{(\mu)}\}$ be a set of Kraus operators for
$\mathcal{E}^r_{B|A}$.  Substituting these and rearranging then gives
\begin{eqnarray}
P(M,N) & = & \sum_{j,k = 1}^{d_A^r} \Bra{k} \tau_A^{\frac{1}{2}} \bm{M}_A \tau_A^{\frac{1}{2}} \Ket{j} \Bra{k} R_{B|A}^{(\mu)\dagger} \bm{N}_B R_{B|A}^{(\mu)} \Ket{j} \\
& = & \sum_{j = 1}^{d_A^r} \Bra{j} (\tau_A^{\frac{1}{2}})^T \bm{M}^T_A (\tau_A^{\frac{1}{2}})^T \sum_{k=1}^{d_A^r} \Ket{k} \Bra{k} R_{B|A}^{(\mu)\dagger} \bm{N}_B R_{B|A}^{(\mu)} \Ket{j}.
\end{eqnarray}
Now, $\tau_A^T = \rho_A$ and $\sum_{k=1}^{d_A^r} \Ket{k} \Bra{k} =P_A$, where $P_A$ is the projector onto the support of $\rho_A$, so
\begin{equation}
P(M,N)  =  \sum_{j = 1}^{d_A^r} \Bra{j} \rho_A^{\frac{1}{2}} \bm{M}^T_A \rho_A^{\frac{1}{2}} P_A R_{B|A}^{(\mu)\dagger} \bm{N}_B R_{B|A}^{(\mu)} \Ket{j}.
\end{equation}
However, $\rho_A^{\frac{1}{2}} P_A R_{B|A}^{(\mu)\dagger} = \rho_A^{\frac{1}{2}} R_{B|A}^{(\mu)\dagger}$, since $R_{B|A}^{(\mu)}$ is only defined on the support of $\rho_A$.  Substituting this and rearranging gives
\begin{equation}
P(M,N) =  \PTr{B}{\bm{N}_B  R_{B|A}^{(\mu)} \sum_{j=1}^{d_A^r} \Ket{j}\Bra{j}_A \rho_A^{\frac{1}{2}} \bm{M}^T_A \rho_A^{\frac{1}{2}} R_{B|A}^{(\mu)\dagger}}.
\end{equation}
Now again $\sum_{j=1}^{d_A^r} \Ket{j}\Bra{j}_A = P_A$ and $ R_{B|A}^{(\mu)} P_A \rho_A^{\frac{1}{2}} =  R_{B|A}^{(\mu)}  \rho_A^{\frac{1}{2}}$, so
\begin{eqnarray}
P(M,N) & = & \PTr{B}{\bm{N}_B \mathcal{E}_{B|A}^r \left ( \rho_A^{\frac{1}{2}} \bm{M}^T_A \rho_A^{\frac{1}{2}} \right )} \\
& = & Q(M,N).
\end{eqnarray}
\end{proof}

\subsection{Commutativity properties of the isomorphism}

\label{New:Commute}

Two commutativity properties of the isomorphism are useful for the applications that follow. Firstly, the isomorphism commutes with the partial trace for tripartite states. To describe this, it is useful to introduce the concept of a reduced map.
\begin{Def}
For a linear map $\mathcal{E}_{BC|A}: \mathfrak{L}(\mathcal{H}_A) \rightarrow \mathfrak{L}(\mathcal{H}_B \otimes \mathcal{H}_C)$.  The reduced map $\mathcal{E}_{B|A}: \mathfrak{L}(\mathcal{H}_A) \rightarrow \mathfrak{L}(\mathcal{H}_B)$ is given by composing the map with the partial trace, i.e. $\mathcal{E}_{B|A} = \text{Tr}_C \circ \mathcal{E}_{BC|A}$.
\end{Def}

Starting with a pair $(\rho_A,\mathcal{E}^r_{BC|A})$, the isomorphism can be used to arrive at a tripartite state $\tau_{ABC}$, and then the partial trace over $C$ gives the bipartite reduced state $\tau_{AB}$.  This is the same bipartite state that one obtains by applying the isomorphism to the pair $(\rho_A,\mathcal{E}^r_{B|A})$.  This is summarized in the following diagram:

\begin{equation}
\label{Cloning:Commute}
\begin{CD}
{\rho_{ABC}} @= {(\rho_A,\mathcal{E}^r_{BC|A})} \\
@V{\text{Tr}_C}VV @VV{\text{Tr}_C}V \\
{\rho_{AB}} @= {(\rho_A,\mathcal{E}^r_{B|A})}.
\end{CD}
\end{equation}

The second commutativity property concerns $M$-measurements.  Starting with a pair $(\rho_A,\mathcal{E}^r_{B|A})$, the isomorphism can be used to arrive at a bipartite state $\tau_{AB}$, and then an $M$-measurement can be applied to system $A$, giving a bipartite state $\sqrt{\bm{M}}_A \otimes I_B \tau_{AB} \sqrt{\bm{M}}_A \otimes I_B$, where the normalization factor has been omitted.  This is the same bipartite state that one obtains by first performing an $M^T$-measurement on $\rho_A$ to obtain the pair $(\sqrt{\bm{M}}^T_A\rho_A \sqrt{\bm{M}}^T_A,\mathcal{E}^r_{B|A})$, and then applying the isomorphism.  This is summarized in the following diagram:

\begin{equation}
\label{Cloning:Commute2}
\begin{CD}
\tau_{AB} @= (\rho_A,\mathcal{E}^r_{B|A}) \\
@V{M_A\text{-measurement}}VV @VV{M_A^T\text{-measurement}}V \\
\sqrt{\bm{M}}_A \otimes I_B \tau_{AB} \sqrt{\bm{M}}_A \otimes I_B @= (\sqrt{\bm{M}}^T_A \rho_A \sqrt{\bm{M}}^T_A , \mathcal{E}^r_{B|A})
\end{CD}
\end{equation}

These commutativity properties are straightforward to prove from the definition of the isomorphism, and so the proofs are omitted here.

\subsection{Remarks}

\label{New:Remarks}

As with the standard isomorphism, the new construction depends on
the basis chosen for $\Ket{\Phi^+}_{AA'}$.  The forward direction
takes a particularly simple form if this is chosen to be an eigenbasis of $\rho_A$,
since this basis is then a Schmidt basis for $\Ket{\Phi}_{AA'}$.  Let $\rho_A = \sum_j \lambda_j \Ket{j}\Bra{j}_A$ be an eigendecomposition of $\rho_A$.  Then eq. (\ref{New:Phi}) can be written as
\begin{equation}
\Ket{\Phi}_{AA'} = \sum_j \sqrt{\lambda_j} \Ket{j}_A \otimes \Ket{j}_{A'},
\end{equation}
and eq. (\ref{New:RhoAB}) reduces to
\begin{equation}
\tau_{AB} = \sum_{jk} \sqrt{\lambda_j \lambda_k} \Ket{j}\Bra{k}_A \otimes \mathcal{E}_{B|A'}^r \left ( \Ket{j}\Bra{k}_{A'} \right ).
\end{equation}
With this choice of basis, the cumbersome transpose can be eliminated,
since $\rho_A^T = \rho_A$.  Additionally,  $\bm{M}^T = \bm{M}$ holds
for any POVM $M$ with operators $\bm{M}$ that are diagonal in this basis.

Note that if $\mathcal{E}_{B|A}$ is a unitary operation, then the
state $\tau_{AB}$ is pure, regardless of the state $\rho_A$.  If in addition $\rho_A$ is of rank $\geq 2$, then $\tau_{AB}$ has more than one Schmidt coefficient, so it is both pure and entangled.

\section{Application: Cloning, Broadcasting and the Monogamy of Entanglement}

\label{Cloning}

The standard Jamio{\l}kowski isomorphism is useful because it allows facts about CP-maps to be recast as facts about bipartite states and vice versa.  On the other hand, there are situations in which it is not necessary to know the action of a TPCP map on the whole Hilbert space, but only how it acts on a particular density matrix, or more
generally on an ensemble decomposition of a particular density matrix.
In such cases, the present isomorphism is a more appropriate tool to
use.  A simple example of this is given by the no-cloning and
no-broadcasting theorems \cite{WZClone82, DieksClone82, NoBroad96}.
In their original form, these theorems concern the possible action of
a TPCP map on just a pair of noncommuting input states.  The
isomorphism allows these theorems to be recast as facts about the
monogamy of entanglement in tripartite states.  These terms are
defined precisely in \S\ref{Cloning:Definitions}.
\S\ref{Cloning:Universal} gives a simple result for the no-universal
broadcasting theorem and \S\ref{Cloning:Ensemble} describes the most general results, including a simplified proof of the no broadcasting theorem that is derived as a byproduct.

Before getting into the technical details, a word of warning about how to interpret the results described below.  The goal is to the relate properties of hypothetical TPCP maps of the form $\mathcal{E}_{BC|A}:\mathfrak{L}(\mathcal{H}_A) \rightarrow \mathfrak{L}(\mathcal{H}_B \otimes \mathcal{H}_C)$ to properties of hypothetical tripartite states $\tau_{ABC}$ via the isomorphism.  The existence of such maps and states is known to be in contradiction with quantum mechanics via the no-cloning/no-broadcasting theorems and the monogamy of entanglement respectively.  Nevertheless, properties of $\tau_{ABC}$ are derived by assuming the existence of $\mathcal{E}_{BC|A}$ and the correctness of quantum mechanics as premises.  This may seem meaningless, since any statement can be logically deduced from a contradiction, regardless of its truth or falsity.  However, the commutativity of the isomorphism with the partial trace averts this conclusion.  As described below, the map $\mathcal{E}_{BC|A}$ is defined by placing constraints on the reduced maps $\mathcal{E}_{B|A}$ and $\mathcal{E}_{C|A}$.  Quantum mechanics does in fact allow maps that satisfy these constraints when they are considered in isolation.  The no-cloning/no-broadcasting theorems simply show that they cannot both be the reduced maps of some global map $\mathcal{E}_{BC|A}$.  Similarly, the reduced states $\tau_{AB}$ and $\tau_{AC}$ of the hypothetical $\tau_{ABC}$ are in fact valid density operators, it is just that they cannot both be the reduced states of some valid global density operator $\tau_{ABC}$.

The results below simply relate the properties of the valid reduced
maps, $\mathcal{E}_{B|A}$ and $\mathcal{E}_{C|A}$, to those of the
valid reduced states, $\rho_{AB}$ and $\rho_{AC}$, via the
isomorphism.  Thereby, no contradiction is involved, and a precise
connection between the properties that make the reduced maps
incompatible with a global map and those that make the reduced states
incompatible with a global state is obtained.

\subsection{Definitions}

\label{Cloning:Definitions}

The definitions of broadcasting and cloning concern TPCP-maps that have an output Hilbert space which is a tensor product of two copies of the input Hilbert space.  With a view to applying the isomorphism, it is useful to continue distinguishing the three copies of the Hilbert space by assigning them different labels $A,A'$ and $A''$.  In this section, states denoted by the same Greek letter, differing only in their subsystem label, refer to the same state on different copies of the same Hilbert space, e.g. $\rho_{A}, \rho_{A'}$ and $\rho_{A''}$ all refer to the same state.  

\begin{Def}
A TPCP-map $\mathcal{E}_{A'A''|A}:\mathfrak{L}(\mathcal{H}_A) \rightarrow \mathfrak{L}(\mathcal{H}_{A'}\otimes\mathcal{H}_{A''})$ is \emph{broadcasting} for a state $\rho_A$ if
\begin{equation}
\mathcal{E}_{A'A''|A}(\rho_A) = \sigma_{A'A''}, \qquad \text{where} \qquad \PTr{A''}{\sigma_{A'A''}} = \rho_{A'} \qquad \text{and} \qquad \PTr{A'}{\sigma_{A'A''}} = \rho_{A''}.
\end{equation}
Equivalently, the reduced maps of $\mathcal{E}_{A'A''|A}$ must satisfy
\begin{equation}
\mathcal{E}_{A'|A}(\rho_A) = \rho_{A'}, \qquad \mathcal{E}_{A''|A}(\rho_A) = \rho_{A''},
\end{equation}
\end{Def}

\begin{Def}
A TPCP-map $\mathcal{E}_{A'A''|A}:\mathfrak{L}(\mathcal{H}_A) \rightarrow \mathfrak{L}(\mathcal{H}_{A'}\otimes\mathcal{H}_{A''})$ is \emph{cloning} for a state $\rho_A$ if 
\begin{equation}
\mathcal{E}_{A'A''|A}(\rho_A) = \rho_{A'} \otimes \rho_{A''}.
\end{equation}
\end{Def}

Cloning is a stronger requirement than broadcasting, since the output
state is required to be a product.  For example, if the input state is
maximally mixed $\rho_A = \frac{1}{d_A} I_A$, then two possible output
states for a broadcasting map are $\Ket{\Phi^+}\Bra{\Phi^+}_{A'A''}$
and $\frac{1}{d_{A'}d_{A''}}I_{A'}\otimes I_{A''}$, but a cloning map
can only output the latter.  In the case of a pure state, cloning and
broadcasting are equivalent, since the purity of the reduced output
states on $A'$ and $A''$ ensures that they must be a product.  Only
the broadcasting condition is needed below, but it is referred to as cloning when only pure states are being considered. 

\begin{Def}
A \emph{universal} broadcasting map is a TPCP map
$\mathcal{E}_{A'A''|A}:\mathfrak{L}(\mathcal{H}_A) \rightarrow
\mathfrak{L}(\mathcal{H}_{A'}\otimes\mathcal{H}_{A''})$ that is
broadcasting for every possible input state.  Equivalently, both the
reduced maps $\mathcal{E}_{A'|A}$ and $\mathcal{E}_{A''|A}$ are the
identity map.
\end{Def}

The original no-cloning theorem \cite{WZClone82, DieksClone82} states that there is no TPCP map that is cloning for a pair of nonorthogonal and nonidentical pure states, and the original no broadcasting theorem \cite{NoBroad96} states that there is no TPCP map that is broadcasting for a pair of noncommuting density operators.  This obviously implies that universal broadcasting is impossible too, but it is worth considering as a special case because the connection between no universal cloning and the monogamy of entanglement is considerably simpler to prove than the general case.

\begin{Def}
A TPCP map is broadcasting for an ensemble of states $\{(p_j,\rho_j)\}$ if it is broadcasting for every state $\rho_j$ in the ensemble.
\end{Def}
Strictly speaking, the weights $p_j$ of the states in the ensemble are irrelevant to the definition, but introducing them is useful for deriving the connection to monogamy of entanglement.  This is because the ensemble average state $\rho_A = \sum_j p_j \rho_j$ can be used along with the reduced maps $\mathcal{E}_{A'|A}$ and $\mathcal{E}_{A''|A}$ to construct bipartite states via the isomorphism.

Note that broadcasting and cloning are often defined in a
superficially more general way than the definitions given here, by
allowing the input to include an arbitrary ancillary system in a
standard state, and the output to also include an ancillary system.
However, the standard theorems about the representations of reduced
dynamics by $CP$-maps \cite{NielChuang} ensure that the present definitions are equivalent.

The monogamy of entanglement refers to the fact that two bipartite states $\rho_{AA'}$ and $\rho_{AA''}$ cannot be arbitrarily entangled if they are the reduced states of a tripartite state $\rho_{AA'A''}$ \cite{Terhal03}.  Typically, there is a tradeoff such that the greater the entanglement of $\rho_{AA'}$, according to some entanglement measure, the lower the entanglement of $\rho_{AA''}$ \cite{CofKunWoo00, DenWoo01, KoaBuzImo00, KoaWin03, OsbVer05}.  For present purposes, it is sufficient to note that if $\rho_{AA'}$ and $\rho_{AA''}$ are both pure, then they must both be product states in order to be compatible with a global state $\rho_{AA'A''}$.

\subsection{Universal Broadcasting}

\label{Cloning:Universal}

The standard Jamio{\l}kowski isomorphism can be used to derive a connection between the no-universal broadcasting theorem, and the monogamy of entanglement. Its proof is much simpler than the more general case described below, so it is included here for completeness.

\begin{Theorem}
Supposing the existence of a universal broadcasting map $\mathcal{E}_{A'A''|A}$ is equivalent to supposing the existence of a tripartite state $\tau_{AA'A''}$, where both the bipartite reduced states $\tau_{AA'}$ and $\tau_{AA''}$ are pure and maximally entangled.
\end{Theorem}

\begin{proof}
By the assumption that $\mathcal{E}_{A'A''|A}$ is universal broadcasting, both the reduced maps $\mathcal{E}_{A'|A}$ and $\mathcal{E}_{A''|A}$ act as the identity on all input states.  The state isomorphic to the identity by the standard isomorphism is the maximally entangled state $\Ket{\Phi^+}$.  Therefore, both $\tau_{AA'}= \Ket{\Phi^+}\Bra{\Phi^+}_{AA'}$ and $\tau_{AA''}= \Ket{\Phi^+}\Bra{\Phi^+}_{AA''}$.

Conversely, assume there is a tripartite state $\tau_{AA'A''}$, such that the reduced states $\tau_{AA'}$ and $\tau_{AA''}$ are pure and maximally entangled.  By acting with independent local unitary transformations on the subsystems $A'$ and $A''$, these states can be transformed to $\Ket{\Phi^+}\Bra{\Phi^+}_{AA'}$ and $\Ket{\Phi^+}\Bra{\Phi^+}_{AA''}$.  The isomorphic maps to these states are both the identity, so the map associated to the transformed tripartite state is universal broadcasting.
\end{proof}

Since this result uses the standard isomorphism, it should not be surprising that the converse has an interpretation in terms of teleportation.  Indeed, if there existed a tripartite state $\tau_{AA'A''}$ with maximally entangled reduced states $\tau_{AA'}$ and $\tau_{AA''}$, then it would be possible to teleport any state to both $A'$ and $A''$ simultaneously, which would provide a method of implementing a universal broadcasting map.

\subsection{Ensemble Broadcasting and Cloning}

\label{Cloning:Ensemble}

In order to generalize this result to the ensemble
broadcasting and cloning, some properties of the fixed point sets of
TPCP-maps are needed. Proofs of the quoted results can be found in
\cite{Paulsen86, Paulsen03}. These are then used to provide a simple
reduction of the no-broadcasting theorem to the no-cloning theorem.
The mathematical structure uncovered in this proof is then used to
derive the connection between ensemble broadcasting and the monogamy
of entanglement.  Finally, this is specialized to pure-state cloning, for
which a stronger result is possible.

\subsubsection{Fixed Point Sets of TPCP-maps}

\label{Cloning:Ensemble:Fixed}
 
The set of density matrices invariant under any TPCP map that acts on $\mathfrak{L}(\mathcal{H})$ is a convex linear subspace of $\mathfrak{L}(\mathcal{H})$.  There is a factorization of the Hilbert space $\mathcal{H}$ into a finite direct sum of tensor products
\begin{equation}
\label{Cloning:Decomp}
\mathcal{H} = \bigoplus_\alpha \mathcal{H}_{\alpha_1} \otimes \mathcal{H}_{\alpha_2},
\end{equation}
such that the invariant density operators are all those of the form
\begin{equation}
\label{Cloning:Inv}
\sum_\alpha q_{\alpha} \mu_{\alpha_1} \otimes \nu_{\alpha_2}
\end{equation} 
where $0 \leq q_\alpha \leq 1, \sum_{\alpha}q_\alpha = 1$.  In this decomposition, the $\mu_{\alpha_1}$'s can be any density operators in $\mathfrak{L}(\mathcal{H}_{\alpha_1})$ and the $\nu_{\alpha_2}$ are fixed density operators in $\mathfrak{L}(\mathcal{H}_{\alpha_2})$.  For a pair of TPCP maps $\mathcal{E}$ and $\mathcal{F}$, the set of density operators invariant under both $\mathcal{E}$ and $\mathcal{F}$ is also of this form.

\subsubsection{The No-Broadcasting Theorem}

\label{Cloning:NoBroadcast}

\begin{Theorem}
A TPCP-map $\mathcal{E}_{A'A''|A}$ that is broadcasting for
a pair of states $\sigma_1, \sigma_2$, where $[\sigma_1,\sigma_2] \neq
0$, is cloning for a set of nonorthogonal and nonidentical pure states.
\end{Theorem}

\begin{proof}
The set of density operators invariant under both the reduced maps $\mathcal{E}_{A'|A}$ and $\mathcal{E}_{A''|A}$ is of the form of eq. (\ref{Cloning:Inv}).  This set must include the density operators $\sigma_1$ and $\sigma_2$, by the assumption that $\mathcal{E}_{A'A''|A}$ is broadcasting, so they can be written as
\begin{equation}
\label{Cloning:EDecomp}
\sigma_j = \sum_\alpha q^{(j)}_{\alpha} \mu^{(j)}_{\alpha_1} \otimes \nu_{\alpha_2}.
\end{equation}
Since $\sigma_1$ and $\sigma_2$ do not commute, there must be at least
one value $\beta$, such that for $\alpha = \beta$, the
$\mathcal{H}_{\beta_1}$ factor in the decomposition of
eq. (\ref{Cloning:Decomp}) is of dimension $\geq 2$,
$[\mu^{(1)}_{\beta_1},\mu^{(2)}_{\beta_1}] \neq 0$ and
$q^{(1)}_{\beta_1},q^{(2)}_{\beta_1} \neq 0$.  Both maps,
$\mathcal{E}_{A'|A}$ and $\mathcal{E}_{A''|A}$, act as the identity on
this factor, and hence any pure state on this factor is cloned by the
map $\mathcal{E}_{A'A''|A}$.  Since the factor is of dimension $\geq
2$, there are nonorthogonal and nonidentical pure states within the factor.   
\end{proof}

This result can be viewed as a simplified proof of the no-broadcasting
theorem, since the no-cloning theorem itself is elementary to prove
\cite{WZClone82, DieksClone82}.  A similar strategy was used by
Lindblad to prove a more general theorem \cite{LindbladClone99}, but the
above is a more direct route to no-broadcasting.

\subsubsection{Ensemble Broadcasting}

\label{Cloning:Ensemble:Ensemble}

\begin{Theorem}
\label{Cloning:Main}
Suppose there existed a TPCP map $\mathcal{E}_{A'A''|A}$, that is broadcasting for a two-element ensemble of states $\{(p, \sigma_1),((1-p),\sigma_2)\}$, such that $[\sigma_1,\sigma_2] \neq 0$.  Let $\rho_A = p \sigma_1 + (1-p) \sigma_2$.  The tripartite state $\tau_{AA'A''}$, isomorphic to $(\rho_A,\mathcal{E}_{A'A''|A}^r)$, would have to be such that it can be transformed with nonzero probability of success into a state that has pure, entangled reduced states on both $AA'$ and $AA''$ by local operations.
\end{Theorem}

\begin{proof}[Proof of theorem \ref{Cloning:Main}]
For the states $\sigma_1, \sigma_2$, use the decomposition given in
eq. (\ref{Cloning:EDecomp}) and define $\beta$ as before.  Now consider the ensemble average density operator $\rho_A = p \sigma_1 + (1-p) \sigma_2$.  This also has a decomposition of the form of eq. (\ref{Cloning:Inv})
\begin{equation}
\rho_A = \sum_\alpha q_{\alpha} \mu_{\alpha_1} \otimes \nu_{\alpha_2}
\end{equation}
where $q_{\alpha} = \Tr{p q^{(1)}_{\alpha} \mu^{(1)}_{\alpha_1} + (1-p) \mu^{(2)}_{\alpha_1}}$ and if $q_{\alpha} \neq 0$ then $\mu_{\alpha_1} = (p q^{(1)}_{\alpha} \mu^{(1)}_{\alpha_1} + (1-p) \mu^{(2)}_{\alpha_1})/q_\alpha$. Now, $\mu_{\beta_1}$ must be of rank $\geq 2$ because $[\mu^{(1)}_{\beta_1},\mu^{(2)}_{\beta_1}] \neq 0$.

Let $P_\alpha$ be the projection operator onto $\mathcal{H}_{\alpha_1} \otimes \mathcal{H}_{\alpha_2}$.  The set of $P_\alpha$ for all values of $\alpha$ is a POVM (in fact it is a Projector Valued Measure) that commutes with $\rho_A$.

Now consider the pair $(\rho_A,\mathcal{E}_{A'|A}^r)$ and construct the isomorphic state $\tau_{AA'}$.  To do this, a basis must be chosen to define the state $\Ket{\Phi^+}$ used to construct the isomorphism.  Choose an eigenbasis of $\rho_A$ to make use of the facts noted in \S\ref{New:Remarks}.

Recall that the state obtained from performing a $P_\alpha$-measurement system $A$ when the state is $\tau_{AA'}$, can be determined by applying the isomorphism to the pair $(P_\alpha \rho_A P_\alpha, \mathcal{E}_{A'|A}^r)$, where the $^T$ is omitted because $P_\alpha^T = P_\alpha$, and the square root is omitted because $P_\alpha$ is idempotent.  Suppose the outcome $\beta$ is obtained, which happens with nonzero probability of success.  Then the updated state after the measurement is
\begin{equation}
P_\beta \rho_A P_\beta = \mu_{\beta_1} \otimes \nu_{\beta_2}.
\end{equation}
Since the map $\mathcal{E}_{A'|A}$ acts as the identity on the factor $\mathcal{H}_{\beta_1}$, the isomorphic state on $AA'$ is of the form $\Ket{\psi}\Bra{\psi} \otimes \eta$, where $\Ket{\psi}$ is a pure state on $\mathcal{H}_{\beta_1} \otimes \mathcal{H}_{\beta_1}$ and $\eta$ is a state on $\mathcal{H}_{\beta_2} \otimes \mathcal{H}_{\beta_2}$.  One copy of $\mathcal{H}_{\beta_1}$ and $\mathcal{H}_{\beta_2}$ belongs to system $A$ and the other belongs to system $A'$.  The state $\Ket{\psi}$ is entangled, since $\rho_{\beta_1}$ is of rank $\geq 2$, and the rank of $\rho_{\beta_1}$ is the number of Schmidt coefficients of $\Ket{\psi}$.

Thus, starting from the state $\tau_{AA'}$, a pure entangled state can be obtained with nonzero probability of success by performing a $P_\alpha$ measurement on system $A$ and discarding the two copies of the subsystem $\mathcal{H}_{\beta_2}$ if the $\beta$ outcome is obtained.  The same argument applies to $\tau_{AA''}$, the state isomorphic to $(\rho_A, \mathcal{E}_{A''|A}^r)$, so we have the desired result.
\end{proof}

\subsubsection{Ensemble Cloning}

For pure state ensemble cloning, a stronger result is possible, which removes the need to perform a measurement on system $A$.
\begin{Theorem}
Suppose there existed a cloning map $\mathcal{E}_{A'A''|A}$ for an ensemble of $\geq 2$ pairwise nonorthogonal, and nonidentical, pure states $\{(p_j,\Ket{\psi_j})\}$, $0 < p_j < 1$, $\sum_j p_j =1$.  Let $\rho_A = \sum_j p_j \Ket{\psi_j}\Bra{\psi_j}$. The tripartite state $\rho_{AA'A''}$ isomorphic to $(\rho_A,\mathcal{E}^r_{A'A''|A})$ would have to be such that both reduced states $\rho_{AA'}$,  $\rho_{AA'}$ are pure and entangled. 
\end{Theorem}

\begin{proof}
Each state $\Ket{\psi_j}\Bra{\psi_j}$ is in the fixed point set of both the reduced maps $\mathcal{E}_{A'|A}$ and $\mathcal{E}_{A''|A}$, and the common fixed point set is of the form of (\ref{Cloning:Inv}).  Therefore, each state $\Ket{\psi_j}$ must be of the form $\Ket{\psi_j} = \Ket{\phi_j}_{\alpha_1} \otimes \Ket{\theta_j}_{\alpha_2}$, where $\Ket{\phi_j}_{\alpha_1}$ is a state on a factor $\mathcal{H}_{\alpha_1}$ and $\Ket{\theta_j}_{\alpha_2}$ is a fixed state on a factor $\mathcal{H}_{\alpha_2}$.  In fact, the factor $\alpha$ must be the same for all the states $\Ket{\psi_j}$, since otherwise they would be orthogonal.  That means that $\Ket{\theta_j}_{\alpha_2}$ must be the same state, $\Ket{\theta}_{\alpha_2}$, for all $j$, and that the $\Ket{\phi_j}_{\alpha_1}$'s are nonorthogonal and nonidentical. The state $\rho_A$ can then be written as $\rho_A = \sum_j \mu_{\alpha_1} \otimes \Ket{\theta}_{\alpha_2}\Bra{\theta}_{\alpha_2}$, where $\mu_{\alpha_1} = \sum_j p_j \Ket{\phi_j}\Bra{\phi_j}_{\alpha_1}$.  Note that $\mu_{\alpha_1}$ is of rank $\geq 2$, and the map $\mathcal{E}_{A'|A}$ acts as the identity on $\mathcal{H}_{\alpha_1}$ and on the state $\Ket{\theta}_{\alpha_2}$.

The isomorphic state $\rho_{AA'}$ is therefore of the form
\begin{equation}
\Ket{\xi}\Bra{\xi} \otimes \Ket{\theta}\Bra{\theta}_{\alpha_2} \otimes \Ket{\theta}\Bra{\theta}_{\alpha_2}
\end{equation}
where $\Ket{\xi}$ is an entangled pure state on $\mathcal{H}_{\alpha_1} \otimes \mathcal{H}_{\alpha_1}$.  One copy of $\mathcal{H}_{\alpha_1}$ belongs to the subsystem $A$ and the other to $A'$, so the state is both pure and entangled.  The same argument applies for the other reduced state $\rho_{AA''}$.
\end{proof}

\section{Discussion}

\label{Discuss}

In this paper, an alternative variant of the Jamio{\l}kowski
isomorphism was derived, and used to demonstrate the connection
between the no-cloning/no-broadcasting theorems and the monogamy of
entanglement.  It is likely that the new isomorphism can be applied in
a variety of other parts of quantum information theory, whenever the
action of a TPCP map on a particular ensemble of states is of
interest, rather than its action on the entire Hilbert space.  For
example, this occurs in prepare-and-measure quantum key distribution
schemes \cite{BB84, B92}.

A possible future project would be to derive bounds on the maximum obtainable fidelity in approximate ensemble broadcasting from the known inequalities for the monogamy of entanglement \cite{CofKunWoo00, DenWoo01, KoaBuzImo00, KoaWin03, OsbVer05}.  It seems plausible that the closer the bipartite reduced states can be made to the ones obtained from the isomorphism, the better the fidelity of the broadcast copies would be.  Fewer results are known about approximate broadcasting for mixed states \cite{DArMacPer05, BusDArMacPer05, DArPerSac06-1, DArPerSac06-2, BusDArMacPer06} than for approximate cloning of pure states, so this could be a fruitful route to pursue.  The main difficulty is that the entanglement measures used in monogamy inequalities are typically not related to fidelity in a straightforward way.  

From a more foundational point of view, we have shown that the map
$\mathcal{E}^r_{B|A}$ mimics the behavior of classical conditional
probability very closely.  The alternative definition of quantum
conditional probability proposed by Cerf and Adami \cite{CerfAdami97,CerfAdami98, CerfAdami99} shares a different set of properties with its classical counterpart, particularly the role of conditional probability in the definition of conditional entropy.  One might ask whether there exists a unified notion of quantum conditional probability that shares all these properties, or whether certain properties of conditional probability are mutually exclusive when raised to the quantum domain.  The answer to this question could be of practical use, since there are several classical probabilistic structures that are usually defined in terms of conditional probabilities, such as Markov Chains and Bayesian Networks \cite{Nea90}.  These might have more than one quantum generalization if the quantum analog of conditional probability is not unique.  

More speculatively, the analogy to conditional probability offers some
hope that a formalism for an abstract quantum probability without any
background causal structures might be obtainable, perhaps within the
framework recently proposed by Hardy \cite{Hardy05}.  One might hope that such a theory would give new insights into how to apply quantum theory to cases in which the background causal structure is unknown a priori, as in quantum gravity.

\begin{acknowledgments}
I would like to thank Howard Barnum, Jonathan Barrett, Lucien Hardy, Nick Jones and Rob Spekkens for useful dicussions.  Part of this work was completed whilst the author was a visitor in Michael Nielsen's research group at the University of Queensland.
\end{acknowledgments}

\bibliography{JamBib}

\begin{thebibliography}{41}
\expandafter\ifx\csname natexlab\endcsname\relax\def\natexlab#1{#1}\fi
\expandafter\ifx\csname bibnamefont\endcsname\relax
  \def\bibnamefont#1{#1}\fi
\expandafter\ifx\csname bibfnamefont\endcsname\relax
  \def\bibfnamefont#1{#1}\fi
\expandafter\ifx\csname citenamefont\endcsname\relax
  \def\citenamefont#1{#1}\fi
\expandafter\ifx\csname url\endcsname\relax
  \def\url#1{\texttt{#1}}\fi
\expandafter\ifx\csname urlprefix\endcsname\relax\def\urlprefix{URL }\fi
\providecommand{\bibinfo}[2]{#2}
\providecommand{\eprint}[2][]{\url{#2}}

\bibitem[{\citenamefont{Nielsen and Chuang}(2000)}]{NielChuang}
\bibinfo{author}{\bibfnamefont{M.~A.} \bibnamefont{Nielsen}} \bibnamefont{and}
  \bibinfo{author}{\bibfnamefont{I.~L.} \bibnamefont{Chuang}},
  \emph{\bibinfo{title}{Quantum Computation and Quantum Information}}
  (\bibinfo{publisher}{Cambridge University Press}, \bibinfo{year}{2000}).

\bibitem[{\citenamefont{Kolmogorov}(1929)}]{Kolmogorov29}
\bibinfo{author}{\bibfnamefont{A.~N.} \bibnamefont{Kolmogorov}},
  \bibinfo{journal}{(in Russian)}  (\bibinfo{year}{1929}).

\bibitem[{\citenamefont{Doob}(1942)}]{Doob42}
\bibinfo{author}{\bibfnamefont{J.~L.} \bibnamefont{Doob}},
  \bibinfo{journal}{Amer. Math. Monthly} \textbf{\bibinfo{volume}{49}},
  \bibinfo{pages}{648} (\bibinfo{year}{1942}).

\bibitem[{\citenamefont{Jamio{\l}kowski}(1972)}]{Jam72}
\bibinfo{author}{\bibfnamefont{A.}~\bibnamefont{Jamio{\l}kowski}},
  \bibinfo{journal}{Rep. Math. Phys.} \textbf{\bibinfo{volume}{3}},
  \bibinfo{pages}{275} (\bibinfo{year}{1972}).

\bibitem[{\citenamefont{Choi}(1975)}]{Choi75}
\bibinfo{author}{\bibfnamefont{M.~D.} \bibnamefont{Choi}},
  \bibinfo{journal}{Lin. Alg. Appl.} \textbf{\bibinfo{volume}{10}},
  \bibinfo{pages}{285} (\bibinfo{year}{1975}).

\bibitem[{\citenamefont{Verstraete and Verschelde}()}]{VerVer03}
\bibinfo{author}{\bibfnamefont{F.}~\bibnamefont{Verstraete}} \bibnamefont{and}
  \bibinfo{author}{\bibfnamefont{H.}~\bibnamefont{Verschelde}},
  \eprint{quant-ph/0202124}.

\bibitem[{\citenamefont{Arrighi and Patricot}(2004)}]{ArrPat04}
\bibinfo{author}{\bibfnamefont{P.}~\bibnamefont{Arrighi}} \bibnamefont{and}
  \bibinfo{author}{\bibfnamefont{C.}~\bibnamefont{Patricot}},
  \bibinfo{journal}{Ann. Phys.} \textbf{\bibinfo{volume}{311}},
  \bibinfo{pages}{26} (\bibinfo{year}{2004}).

\bibitem[{\citenamefont{Wootters and Zurek}(1982)}]{WZClone82}
\bibinfo{author}{\bibfnamefont{W.~K.} \bibnamefont{Wootters}} \bibnamefont{and}
  \bibinfo{author}{\bibfnamefont{W.~H.} \bibnamefont{Zurek}},
  \bibinfo{journal}{Nature} \textbf{\bibinfo{volume}{299}},
  \bibinfo{pages}{802} (\bibinfo{year}{1982}).

\bibitem[{\citenamefont{Dieks}(1982)}]{DieksClone82}
\bibinfo{author}{\bibfnamefont{D.}~\bibnamefont{Dieks}},
  \bibinfo{journal}{Phys. Lett.} \textbf{\bibinfo{volume}{92A}},
  \bibinfo{pages}{271} (\bibinfo{year}{1982}).

\bibitem[{\citenamefont{Barnum et~al.}(1996)\citenamefont{Barnum, Caves, Fuchs,
  Jozsa, and Schumacher}}]{NoBroad96}
\bibinfo{author}{\bibfnamefont{H.}~\bibnamefont{Barnum}},
  \bibinfo{author}{\bibfnamefont{C.~M.} \bibnamefont{Caves}},
  \bibinfo{author}{\bibfnamefont{C.~A.} \bibnamefont{Fuchs}},
  \bibinfo{author}{\bibfnamefont{R.}~\bibnamefont{Jozsa}}, \bibnamefont{and}
  \bibinfo{author}{\bibfnamefont{B.}~\bibnamefont{Schumacher}},
  \bibinfo{journal}{Phys. Rev. Lett.} \textbf{\bibinfo{volume}{79}},
  \bibinfo{pages}{2818} (\bibinfo{year}{1996}), \eprint{quant-ph/9511010}.

\bibitem[{\citenamefont{Terhal}(2004)}]{Terhal03}
\bibinfo{author}{\bibfnamefont{B.~M.} \bibnamefont{Terhal}},
  \bibinfo{journal}{IBM Journal of Research and Development}
  \textbf{\bibinfo{volume}{48}}, \bibinfo{pages}{71} (\bibinfo{year}{2004}),
  \eprint{quant-ph/0307120}.

\bibitem[{\citenamefont{Ohya}(1983{\natexlab{a}})}]{Ohya83-1}
\bibinfo{author}{\bibfnamefont{M.}~\bibnamefont{Ohya}}, \bibinfo{journal}{IEEE
  T. Inform. Theory} \textbf{\bibinfo{volume}{29}}, \bibinfo{pages}{770}
  (\bibinfo{year}{1983}{\natexlab{a}}).

\bibitem[{\citenamefont{Ohya}(1983{\natexlab{b}})}]{Ohya83-2}
\bibinfo{author}{\bibfnamefont{M.}~\bibnamefont{Ohya}}, \bibinfo{journal}{Lett.
  Nuovo Cimento} \textbf{\bibinfo{volume}{38}}, \bibinfo{pages}{402}
  (\bibinfo{year}{1983}{\natexlab{b}}).

\bibitem[{\citenamefont{Griffiths}(2005)}]{Griffiths05}
\bibinfo{author}{\bibfnamefont{R.~B.} \bibnamefont{Griffiths}},
  \bibinfo{journal}{Phys. Rev. A} \textbf{\bibinfo{volume}{71}},
  \bibinfo{pages}{042337} (\bibinfo{year}{2005}), \eprint{quant-ph/0409106}.

\bibitem[{\citenamefont{D'Ariano and Lo~Presti}(2003)}]{LoPresti03}
\bibinfo{author}{\bibfnamefont{G.~M.} \bibnamefont{D'Ariano}} \bibnamefont{and}
  \bibinfo{author}{\bibfnamefont{P.}~\bibnamefont{Lo~Presti}},
  \bibinfo{journal}{Phys. Rev. Lett.} \textbf{\bibinfo{volume}{91}},
  \bibinfo{pages}{047902} (\bibinfo{year}{2003}), \eprint{quant-ph/0211133}.

\bibitem[{\citenamefont{Fuchs}(2002)}]{Fuchs02}
\bibinfo{author}{\bibfnamefont{C.~A.} \bibnamefont{Fuchs}}
  (\bibinfo{year}{2002}), \eprint{quant-ph/0205039}.

\bibitem[{\citenamefont{Fuchs}(2003)}]{Fuchs03}
\bibinfo{author}{\bibfnamefont{C.~A.} \bibnamefont{Fuchs}},
  \bibinfo{journal}{J. Mod. Opt.} \textbf{\bibinfo{volume}{50}},
  \bibinfo{pages}{987} (\bibinfo{year}{2003}).

\bibitem[{\citenamefont{Asorey et~al.}(2005)\citenamefont{Asorey, Kossakowski,
  Marmo, and Sudarshan}}]{AsoreyEtAl05}
\bibinfo{author}{\bibfnamefont{M.}~\bibnamefont{Asorey}},
  \bibinfo{author}{\bibfnamefont{A.}~\bibnamefont{Kossakowski}},
  \bibinfo{author}{\bibfnamefont{G.}~\bibnamefont{Marmo}}, \bibnamefont{and}
  \bibinfo{author}{\bibfnamefont{E.~C.~G.} \bibnamefont{Sudarshan}},
  \bibinfo{journal}{Open Syst. Inf. Dyn.} \textbf{\bibinfo{volume}{12}},
  \bibinfo{pages}{319} (\bibinfo{year}{2005}), \eprint{quan-ph/0602228}.

\bibitem[{\citenamefont{Cerf and Adami}(1997)}]{CerfAdami97}
\bibinfo{author}{\bibfnamefont{N.~J.} \bibnamefont{Cerf}} \bibnamefont{and}
  \bibinfo{author}{\bibfnamefont{C.}~\bibnamefont{Adami}},
  \bibinfo{journal}{Phys. Rev. Lett.} \textbf{\bibinfo{volume}{79}},
  \bibinfo{pages}{5194} (\bibinfo{year}{1997}), \eprint{quant-ph/9512022}.

\bibitem[{\citenamefont{Cerf and Adami}(1998)}]{CerfAdami98}
\bibinfo{author}{\bibfnamefont{N.~J.} \bibnamefont{Cerf}} \bibnamefont{and}
  \bibinfo{author}{\bibfnamefont{C.}~\bibnamefont{Adami}},
  \bibinfo{journal}{Physica D} \textbf{\bibinfo{volume}{120}},
  \bibinfo{pages}{62} (\bibinfo{year}{1998}), \eprint{quant-ph/9605039}.

\bibitem[{\citenamefont{Cerf and Adami}(1999)}]{CerfAdami99}
\bibinfo{author}{\bibfnamefont{N.~J.} \bibnamefont{Cerf}} \bibnamefont{and}
  \bibinfo{author}{\bibfnamefont{C.}~\bibnamefont{Adami}},
  \bibinfo{journal}{Phys. Rev. A} \textbf{\bibinfo{volume}{60}},
  \bibinfo{pages}{893} (\bibinfo{year}{1999}).

\bibitem[{\citenamefont{{\.Z}yczkowski and Bengtsson}(2004)}]{ZyczBeng04}
\bibinfo{author}{\bibfnamefont{K.}~\bibnamefont{{\.Z}yczkowski}}
  \bibnamefont{and}
  \bibinfo{author}{\bibfnamefont{I.}~\bibnamefont{Bengtsson}},
  \bibinfo{journal}{Open Syst. Inf. Dyn.} \textbf{\bibinfo{volume}{11}},
  \bibinfo{pages}{3} (\bibinfo{year}{2004}), \eprint{quant-ph/0401119}.

\bibitem[{\citenamefont{Nielsen and Chuang}(1997)}]{NielChuang97}
\bibinfo{author}{\bibfnamefont{M.~A.} \bibnamefont{Nielsen}} \bibnamefont{and}
  \bibinfo{author}{\bibfnamefont{I.~L.} \bibnamefont{Chuang}},
  \bibinfo{journal}{Phys. Rev. Lett.} \textbf{\bibinfo{volume}{79}},
  \bibinfo{pages}{321} (\bibinfo{year}{1997}), \eprint{quant-ph/9703032}.

\bibitem[{\citenamefont{L{\"u}ders}(1951)}]{Luders}
\bibinfo{author}{\bibfnamefont{G.}~\bibnamefont{L{\"u}ders}},
  \bibinfo{journal}{Ann. Physik} \textbf{\bibinfo{volume}{8}},
  \bibinfo{pages}{322} (\bibinfo{year}{1951}), \bibinfo{note}{translated by K.
  A. Kirkpatrick, quant-ph/0403007}.

\bibitem[{\citenamefont{Coffman et~al.}(2000)\citenamefont{Coffman, Kundu, and
  Wootters}}]{CofKunWoo00}
\bibinfo{author}{\bibfnamefont{V.}~\bibnamefont{Coffman}},
  \bibinfo{author}{\bibfnamefont{J.}~\bibnamefont{Kundu}}, \bibnamefont{and}
  \bibinfo{author}{\bibfnamefont{W.~K.} \bibnamefont{Wootters}},
  \bibinfo{journal}{Phys. Rev. A} \textbf{\bibinfo{volume}{61}},
  \bibinfo{pages}{052306} (\bibinfo{year}{2000}), \eprint{quant-ph/9907047}.

\bibitem[{\citenamefont{Dennison and Wootters}(2002)}]{DenWoo01}
\bibinfo{author}{\bibfnamefont{K.~A.} \bibnamefont{Dennison}} \bibnamefont{and}
  \bibinfo{author}{\bibfnamefont{W.~K.} \bibnamefont{Wootters}},
  \bibinfo{journal}{Phys. Rev. A} \textbf{\bibinfo{volume}{65}},
  \bibinfo{pages}{010301} (\bibinfo{year}{2002}), \eprint{quant-ph/0106058}.

\bibitem[{\citenamefont{Koashi et~al.}(2000)\citenamefont{Koashi, Bu{\v{z}}ek,
  and Imoto}}]{KoaBuzImo00}
\bibinfo{author}{\bibfnamefont{M.}~\bibnamefont{Koashi}},
  \bibinfo{author}{\bibfnamefont{V.}~\bibnamefont{Bu{\v{z}}ek}},
  \bibnamefont{and} \bibinfo{author}{\bibfnamefont{N.}~\bibnamefont{Imoto}},
  \bibinfo{journal}{Phys. Rev. A} \textbf{\bibinfo{volume}{62}},
  \bibinfo{pages}{050302} (\bibinfo{year}{2000}), \eprint{quant-ph/0007086}.

\bibitem[{\citenamefont{Koashi and Winter}(2004)}]{KoaWin03}
\bibinfo{author}{\bibfnamefont{M.}~\bibnamefont{Koashi}} \bibnamefont{and}
  \bibinfo{author}{\bibfnamefont{A.}~\bibnamefont{Winter}},
  \bibinfo{journal}{Phys. Rev. A} \textbf{\bibinfo{volume}{69}},
  \bibinfo{pages}{022309} (\bibinfo{year}{2004}), \eprint{quant-ph/0310037}.

\bibitem[{\citenamefont{Osborne and Verstraete}(2005)}]{OsbVer05}
\bibinfo{author}{\bibfnamefont{T.~J.} \bibnamefont{Osborne}} \bibnamefont{and}
  \bibinfo{author}{\bibfnamefont{F.}~\bibnamefont{Verstraete}}
  (\bibinfo{year}{2005}), \eprint{quant-ph/0502176}.

\bibitem[{\citenamefont{Paulsen}(1986)}]{Paulsen86}
\bibinfo{author}{\bibfnamefont{V.~I.} \bibnamefont{Paulsen}},
  \emph{\bibinfo{title}{Completely Bounded Maps and Dilations}}
  (\bibinfo{publisher}{Longman}, \bibinfo{year}{1986}).

\bibitem[{\citenamefont{Paulsen}(2003)}]{Paulsen03}
\bibinfo{author}{\bibfnamefont{V.~I.} \bibnamefont{Paulsen}},
  \emph{\bibinfo{title}{Completely Bounded Maps and Operator Algebras}},
  vol.~\bibinfo{volume}{78} of \emph{\bibinfo{series}{Cambridge Studies in
  Advanced Mathematics}} (\bibinfo{publisher}{Cambridge University Press},
  \bibinfo{year}{2003}).

\bibitem[{\citenamefont{Lindblad}(1999)}]{LindbladClone99}
\bibinfo{author}{\bibfnamefont{G.}~\bibnamefont{Lindblad}},
  \bibinfo{journal}{Lett. Math. Phys.} \textbf{\bibinfo{volume}{47}},
  \bibinfo{pages}{189} (\bibinfo{year}{1999}).

\bibitem[{\citenamefont{Bennett and Brassard}(1984)}]{BB84}
\bibinfo{author}{\bibfnamefont{C.~H.} \bibnamefont{Bennett}} \bibnamefont{and}
  \bibinfo{author}{\bibfnamefont{G.}~\bibnamefont{Brassard}}, in
  \emph{\bibinfo{booktitle}{Proc. IEEE International Conference on Computer
  Systems and Signal Processing}} (\bibinfo{year}{1984}), pp.
  \bibinfo{pages}{175--179}.

\bibitem[{\citenamefont{Bennett}(1992)}]{B92}
\bibinfo{author}{\bibfnamefont{C.~H.} \bibnamefont{Bennett}},
  \bibinfo{journal}{Phys. Rev. Lett.} \textbf{\bibinfo{volume}{68}},
  \bibinfo{pages}{3121} (\bibinfo{year}{1992}).

\bibitem[{\citenamefont{D'Ariano et~al.}(2005)\citenamefont{D'Ariano,
  Macchiavello, and Perinotti}}]{DArMacPer05}
\bibinfo{author}{\bibfnamefont{G.~M.} \bibnamefont{D'Ariano}},
  \bibinfo{author}{\bibfnamefont{C.}~\bibnamefont{Macchiavello}},
  \bibnamefont{and}
  \bibinfo{author}{\bibfnamefont{P.}~\bibnamefont{Perinotti}},
  \bibinfo{journal}{Phys. Rev. Lett.} \textbf{\bibinfo{volume}{95}},
  \bibinfo{pages}{060503} (\bibinfo{year}{2005}), \eprint{quant-ph/0506251}.

\bibitem[{\citenamefont{Buscemi et~al.}(2005)\citenamefont{Buscemi, D'Ariano,
  Macchiavello, and Perinotti}}]{BusDArMacPer05}
\bibinfo{author}{\bibfnamefont{F.}~\bibnamefont{Buscemi}},
  \bibinfo{author}{\bibfnamefont{G.~M.} \bibnamefont{D'Ariano}},
  \bibinfo{author}{\bibfnamefont{C.}~\bibnamefont{Macchiavello}},
  \bibnamefont{and}
  \bibinfo{author}{\bibfnamefont{P.}~\bibnamefont{Perinotti}}, in
  \emph{\bibinfo{booktitle}{13th Quantum Information Technology Symposium
  (QIT13), Tokohu University, Sendal}} (\bibinfo{year}{2005}), pp.
  \bibinfo{pages}{149--155}, \eprint{quant-ph/0510155}.

\bibitem[{\citenamefont{D'Ariano
  et~al.}(2006{\natexlab{a}})\citenamefont{D'Ariano, Perinotti, and
  Sacchi}}]{DArPerSac06-1}
\bibinfo{author}{\bibfnamefont{G.~M.} \bibnamefont{D'Ariano}},
  \bibinfo{author}{\bibfnamefont{P.}~\bibnamefont{Perinotti}},
  \bibnamefont{and} \bibinfo{author}{\bibfnamefont{M.~F.} \bibnamefont{Sacchi}}
  (\bibinfo{year}{2006}{\natexlab{a}}), \eprint{quant-ph/0601114}.

\bibitem[{\citenamefont{D'Ariano
  et~al.}(2006{\natexlab{b}})\citenamefont{D'Ariano, Perinotti, and
  Sacchi}}]{DArPerSac06-2}
\bibinfo{author}{\bibfnamefont{G.~M.} \bibnamefont{D'Ariano}},
  \bibinfo{author}{\bibfnamefont{P.}~\bibnamefont{Perinotti}},
  \bibnamefont{and} \bibinfo{author}{\bibfnamefont{M.~F.} \bibnamefont{Sacchi}}
  (\bibinfo{year}{2006}{\natexlab{b}}), \eprint{quant-ph/0602037}.

\bibitem[{\citenamefont{Buscemi et~al.}(2006)\citenamefont{Buscemi, D'Ariano,
  Macchiavello, and Perinotti}}]{BusDArMacPer06}
\bibinfo{author}{\bibfnamefont{F.}~\bibnamefont{Buscemi}},
  \bibinfo{author}{\bibfnamefont{G.~M.} \bibnamefont{D'Ariano}},
  \bibinfo{author}{\bibfnamefont{C.}~\bibnamefont{Macchiavello}},
  \bibnamefont{and} \bibinfo{author}{\bibfnamefont{P.}~\bibnamefont{Perinotti}}
  (\bibinfo{year}{2006}), \eprint{quant-ph/0602125}.

\bibitem[{\citenamefont{Neapolitan}(1990)}]{Nea90}
\bibinfo{author}{\bibfnamefont{R.~E.} \bibnamefont{Neapolitan}},
  \emph{\bibinfo{title}{Probabilistic Reasoning In Expert Systems}}
  (\bibinfo{publisher}{Wiley}, \bibinfo{year}{1990}).

\bibitem[{\citenamefont{Hardy}(2005)}]{Hardy05}
\bibinfo{author}{\bibfnamefont{L.}~\bibnamefont{Hardy}} (\bibinfo{year}{2005}),
  \eprint{gr-qc/0509120}.

\end{thebibliography}

\end{document}